\documentclass[12pt,a4paper]{article}
  \usepackage[a4paper, left=1.9cm, right=1.9cm, top=1.12cm, bottom=2.7cm]{geometry}
 \usepackage[final]{pdfpages}

\usepackage{amssymb,amsmath,latexsym,amsfonts,mathrsfs}

\usepackage[active]{srcltx}  % SRC Specials: DVI [Inverse] Search
\usepackage[cp1251]{inputenc}
\usepackage[english]{babel}
\usepackage{setspace}
\usepackage{natbib}
\usepackage{textcomp}
\usepackage{dsfont}
\usepackage{graphics}
\usepackage{graphicx}
\usepackage{makeidx}
\usepackage{amsbsy}
\usepackage{verbatim}
\usepackage{authblk}
\usepackage{bm}
\usepackage{bbm}
\usepackage{rotating}
\usepackage{adjustbox}
\usepackage[font=small,labelfont=bf]{caption}
\usepackage{caption}
\usepackage{subfloat}
\usepackage{amsmath}
\usepackage{pdflscape}
\usepackage{geometry}
\usepackage{ulem}
\usepackage{array}

\makeindex

\bibpunct{(}{)}{,}{a}{,}{,}
 \makeatletter
 	
 	\newcommand{\thickhline}{%
 		\noalign {\ifnum 0=`}\fi \hrule height 1pt
 		\futurelet \reserved@a \@xhline
 	}
 \makeatother

 	\newtheorem{thm}{Theorem}
 	\newtheorem{theorem}[thm]{Theorem}
 	\newtheorem{lemma}{Lemma}
 	
    \newtheorem{remark}{Remark}

 	\newenvironment{proofof}[1]{\smallskip\noindent{\it #1 \ }\rm}
 	{\hspace*{\fill} $\Box$ \medskip}

 	\newenvironment{theorem*}[1]{\smallskip\noindent{\bf #1 \ }\rm}
 	{\medskip}

\usepackage{color}

\usepackage{ulem}

\begin{document}
\date{}
\title{Statistical inference for the  EU portfolio in high dimensions}
\author[a]{{Taras Bodnar}}
\author[b]{{Solomiia Dmytriv}}
\author[c]{{Yarema Okhrin}}
\author[d]{Nestor Parolya\thanks{Corresponding Author: Nestor Parolya. E-Mail: N.Parolya@tudelft.nl}}
\author[b]{{Wolfgang Schmid}}

\affil[a]{{\footnotesize \textit{Department of Mathematics, Stockholm University, Stockholm, Sweden}}}
\affil[b]{{\footnotesize \textit{Department of Statistics, European University Viadrina, Frankfurt(Oder), Germany}}}
\affil[c]{{\footnotesize\textit{ Department of Statistics, University of Augsburg, Augsburg, Germany}}}
\affil[d]{{\footnotesize \textit{Delft Institute of Applied Mathematics, Delft University of Technology, The Netherlands}}}
\maketitle
\vspace{-0.7cm}
\begin{abstract}  In this paper, using the shrinkage-based approach for portfolio weights and modern results from random matrix theory we construct an effective procedure for testing the efficiency of the expected utility (EU) portfolio and discuss the asymptotic behavior of the proposed test statistic under the high-dimensional asymptotic regime, namely when the number of assets $p$ increases at the same rate as the sample size $n$ such that their ratio $p/n$ approaches a positive constant  $c\in(0,1)$ as  $n\to\infty$.  We provide an extensive simulation study where the power function and receiver operating characteristic curves of the test are analyzed. In the empirical study, the methodology is applied to the returns of S\&P 500 constituents.
\end{abstract}


\section{Introduction}\label{sec1}
Following the mean-variance approach of \citet{markowitz1952portfolio}, which is considered to be one of the most popular portfolio choice strategies, the weights of an optimal portfolio are obtained by minimizing the portfolio variance for a predefined level of the portfolio expected return. This set of optimal portfolios determines the efficient frontier in the mean-variance space. The Markowitz approach formalizes the advantages of portfolio diversification and has become a benchmark for both researchers and practitioners in portfolio management.

Markowitz optimal portfolios, also known as mean-variance optimal portfolios, can also be obtained as solutions of other optimization problems (e.g., \citet{BODNAR2013637}), like by maximizing the expected quadratic utility (EU) function (see, \citet{ingersoll1987theory}) expressed as
\begin{equation}\label{EU_problem}
\mathbf{w}'\boldsymbol{\mu} - \frac{\gamma}{2} \mathbf{w}'\mathbf{\Sigma} \mathbf{w} \to max \quad\mbox{subject to}\quad \mathbf{w}'\mathbf{1}_p=1,
\end{equation}
where $\mathbf{w}=(w_1, \ldots, w_p)'$ is the vector of portfolio weights, $\mathbf{1}_p$ is the $p$-dimensional vectors of ones, $\boldsymbol{\mu} $ and $\mathbf{\Sigma}$ are the mean vector and the covariance matrix of the random vector of asset returns $\mathbf{x}=(x_1, \ldots, x_p)'$. The quantity $\gamma>0$ measures the investors attitude towards risk. If $\gamma=\infty$, then the investor is fully risk averse and determines the investment strategy by minimizing the portfolio variance without paying attention to the  expected portfolio return, i.e., constructs the so-called global minimum variance (GMV) portfolio. Under the assumption that the asset returns are normally distributed, the problem of maximization the mean-variance objective function (\ref{EU_problem}) is equivalent to the maximization of the expected exponential utility, which implies constant absolute risk aversion (CARA). In this case, $\gamma$ is equal to the investors absolute risk aversion coefficient (see, e.g., \citet{ingersoll1987theory}).

We denote the solution of (\ref{EU_problem}) by $\mathbf{w}_{EU}$ and it is given by
\begin{equation}\label{solution_EU}
\mathbf{w}_{EU}= \frac{\mathbf{\Sigma}^{-1}\textbf{1}_p}{\textbf{1}_p'\mathbf{\Sigma}^{-1}\mathbf{1}_p}+ \gamma^{-1}\mathbf{Q}\boldsymbol{\mu},
\end{equation}
where
\begin{equation}
\mathbf{Q}=\mathbf{\Sigma}^{-1}-\frac{\mathbf{\Sigma}^{-1}\mathbf{1}_p\textbf{1}_p'\mathbf{\Sigma}^{-1}}{\textbf{1}_p'\mathbf{\Sigma}^{-1}\mathbf{1}_p}.
\end{equation}
The case of fully risk averse investor, i.e., $\gamma = \infty$, leads to the weights of the GMV portfolio expressed as
\begin{equation}\label{solution_GMV}
\mathbf{w}_{GMV}= \frac{\mathbf{\Sigma}^{-1}\textbf{1}_p}{\textbf{1}_p'\mathbf{\Sigma}^{-1}\mathbf{1}_p}.
\end{equation}

The derived formulas of optimal portfolio weights \eqref{solution_EU} and \eqref{solution_GMV} cannot directly be used in practice, since they both depend on unknown parameters of the data generating process. The mean vector $\boldsymbol{\mu}$ and the covariance matrix $\mathbf{\Sigma}$ are not observable in practice   and have to be estimated by using historical data for asset returns. This, however, introduces further sources of risk into the investment process, namely the estimation risk which has been ignored for a long time in finance.

The most commonly used approach to estimate the weights of optimal portfolios is based on simple replacing the unknown first two moments of the asset returns by their sample counterparts. As a result, we obtain a ''plug-in'' estimator for the optimal portfolio weights also known as its sample estimator, which is a traditional way to construct a portfolio in practice. Assuming that the asset returns are independent and normally distributed \citet{okhrin2006distributional} obtain the asymptotic distribution of the sample estimator of the EU portfolio weights, while the corresponding exact distributional results can be found in \citet{bodnarEU}. Further theoretical and practically relevant findings related to the characterization of the distribution of the sample estimator of the optimal portfolio weights and their characteristics can be found in \citet{yang2015robust}, \citet{woodgate2015much}, \citet{simaan2018estimation}, \citet{zhao2019optimal}, among others.

The use of the ''plug-in'' estimators in practice has been widely criticized in statistical and financial literature. One of the main drawbacks of the sample estimators is the investors overoptimism about the optimality of the constructed portfolio. Several studies (see, e.g., \citet{siegel2007performance}, \citet{kan2008distribution}, \citet{bodnar2010unbiased}) show with theoretical and empirical arguments that the plug-in estimator of the efficient frontier overestimates the location of the true efficient frontier in the mean-variance space. This leads to too optimistic trading strategies which perform in practice typically much worse than expected.

In recent years, other types of estimators for the optimal portfolio weights have been introduced in the literature. Some estimators attempt to improve the estimators for the parameters of the asset returns. Relying on the idea of \citet{stein1956} we can use a shrinkage estimator for the mean vector and for the covariance matrix or its inverse (see, e.g., \citet{BodnarGuptaParolya2014} and \citet{BodnarGuptaParolya2016}). Alternatively, one can apply the shrinkage method directly to portfolio weights as suggested by \citet{golosnoy2007multivariate}, \citet{okhrin2008estimation}, \citet{frahm2010dominating}, etc. The goal of the approach is to reduce the estimation uncertainty and to decrease the variance in the estimated portfolio weights.

The problem of assessing the estimation risk, when an optimal portfolio is constructed, becomes very challenging from the high-dimensional perspectives, i.e., when both the number of included assets $p$ and the sample size $n$ tend to infinity simultaneously such that $p/n$ tends to the concentration ratio $c>0$ as $n\to\infty$ (\citet{bai2011estimating}). Under the classical asymptotic regime, when the number of assets $p$ is fixed and substantially smaller then the sample size $n$, the traditional ''plug-in'' estimator of optimal portfolio weights is consistent (see, \citet{okhrin2006distributional}, \citet{bodnarEU}). On the other hand, the sample estimators of the mean vector and of the covariance matrix are not longer feasible under the high-dimensional asymptotics (\citet{bai2010spectral}, \citet{bai2011estimating}, \citet{bodnar2019optimal}), which has a negative impact on the performance of the asset allocation strategy. Moreover, the inverse covariance matrix does not exist anymore for $c>1$ and the optimal portfolios cannot be constructed in a traditional way.

Nowadays, the technological advances and the availability of financial information make the whole universe of assets easily accessible for private and institutional investors. This leads to portfolios consisting of hundreds of assets and to a high demand for new results on constructing optimal portfolios in a high-dimensional setting. Similarly as in the low-dimensional case, the first line of the research deals with deriving  improved estimators for the mean vector and the covariance matrix of asset returns. These are used to obtain improved plug-in estimators  of the optimal portfolio weights (see, \citet{ledoit2017nonlinear}, \citet{holgersson2020risk}). The second possibility is to improve the estimators of the optimal portfolio weights directly. This can be achieved  by taking their functional dependence on the mean vector and of  covariance matrix. Following this approach \citet{bodnar2018estimation} suggest the optimal shrinkage estimator for the GMV portfolio weights, while \citet{bodnar2016okhrin} propose the optimal shrinkage estimator for the EU portfolio weights. Both estimators are derived by using recent results in random matrix theory and appear to be feasible even in the case of $c>1$. Other optimal portfolio choice strategies under the high-dimensional regime were established by \citet{rubio2012performance}, \citet{benidis2018sparse}, \citet{zhao2019optimal}.

It is important to note that the statistical methods developed for estimating optimal portfolio weights can be linked to the classical methods used in statistical signal processing. For example, the Capon or minimum variance spatial filter is equivalent to the GMV portfolio in signal processing literature (see \citet{verdu1998} and \citet{vantrees2002}). The estimation risk of the high-dimensional minimum variance beamformer is studied in \citet{rubio2012performance} and \citet{yang2018high}, while its constrained versions are discussed in \citet{LiStoicaWang2004}. Moreover, \citet{MestreLaugunas2006} discuss the finite-sample size effect on minimum variance filter and \citet{Palomar2013} present an improved calibration of the precision matrix. Further literature on the applications of random matrix theory to signal processing and portfolio optimization can be found in \citet{PalomarBook2016} and references therein.

We contribute to the recent literature in  portfolio theory and  signal processing theory by developing new statistical tests on the weights of the EU portfolio in a high-dimensional setting. From practical point of view an investor will have an opportunity to test if the current large portfolio coincides with a prespecified benchmark portfolio or there are significant deviations.  From the theoretical perspective we contribute by derivation of confidence intervals and test theory for expressions including functions of both the mean vector and the covariance matrix. This directly  extends the existent results on testing the structure of the covariance matrix in high-dimensional settings (see, e.g., \citet{bai2009corrections}, \citet{yao2015sample}, \citet{bodnar2019testing}).
The new approach is based on the shrinkage estimator of the EU portfolio weights and extends the one derived for the weights of the GMV portfolio in \citet{BodnarDmytriv2019} by taking the uncertainty about the estimated mean vector into account when the high-dimensional asymptotic distribution of the test statistic is derived. One of the main advantages of the approach is that the whole high-dimensional vector of portfolio weights can be tested in a single step. Moreover, the investor can make a decision about the efficiency of the holding portfolio based on the result of the testing procedure.

The rest of paper is organized as follows. In Section~\ref{sec2}, we describe the existent approaches in testing the finite number of the EU portfolio weights in both low and high dimensions. New test based on the shrinkage approach is suggested in Section~\ref{sec3}. Here, the asymptotic distribution of the test statistic is derived under both the null and the alternative hypotheses under high-dimensional settings. In Section~\ref{sec4a}, we compare the new test with the existent approaches in terms of size and power properties, while an empirical illustration is provided in Section~\ref{sec4b}. Concluding remarks are presented in Section~\label{sec5}.

\section{Sample estimator of the EU portfolio and test theory}\label{sec2}

We consider a financial market consisting of $p$ risky assets. Let $\textbf{x}_t$ denote the $p$-dimensional vector of the returns on risky assets at time $t$. Suppose that $E(\mathbf{x}_t)=\boldsymbol{\mu}$ and $Cov(\mathbf{x}_t)=\mathbf{\Sigma}$ where $\mathbf{\Sigma}$ is assumed to be positive definite. Let $\mathbf{x}_1, \mathbf{x}_2, \ldots,\mathbf{x}_n$ be a sample of asset return vectors consisting of their $n$ independent realizations and let $\mathbf{X}_n=(\mathbf{x}_1, \mathbf{x}_2, \ldots,\mathbf{x}_n)$ stand for the $p \times n$ data matrix. Throughout of the paper we assume that the asset returns are independent and identically normally distributed, i.e. $\mathbf{x}_i\sim \mathcal{N}_p(\boldsymbol{\mu},\mathbf{\Sigma}), \, i=1,\ldots,n$.

The sample estimators of $\boldsymbol{\mu}$ and $\mathbf{\Sigma}$ are given by
\begin{equation}\label{plug-in}
\mathbf{\bar{x}}_{n} = \frac{1}{n}\sum_{j=1}^{n}\mathbf{x}_{j} \,\,\, \mbox{and} \,\,\,
\hat{\mathbf{\Sigma}}_{n}= \frac{1}{n-1}\sum_{j=1}^{n}\left(\mathbf{x}_{j}-\mathbf{\bar{x}}_{n}\right)\left(\textbf{x}_{j}-\mathbf{\bar{x}}_{n}\right)'.
\end{equation}
Replacing $\boldsymbol{\mu}$ and $\mathbf{\Sigma}$ in (\ref{solution_EU}) by their sample estimators from \eqref{plug-in}, we obtain the sample estimator of the EU portfolio weights expressed as
\begin{equation*}\label{solution_EU_est}
\hat{\mathbf{w}}_{EU}= \frac{\hat{\mathbf{\Sigma}}_n^{-1}\textbf{1}_p}{\textbf{1}_p'\hat{\mathbf{\Sigma}}_n^{-1}\mathbf{1}_p}+ \gamma^{-1}\hat{\mathbf{Q}}_n\hat{\boldsymbol{\mu}}_n,
\end{equation*}
where
\begin{equation} \hat{\mathbf{Q}}_n=\hat{\mathbf{\Sigma}}_n^{-1}-\frac{\hat{\mathbf{\Sigma}}_n^{-1}\mathbf{1}_p\textbf{1}_p'\hat{\mathbf{\Sigma}}_n^{-1}}{\textbf{1}_p'\hat{\mathbf{\Sigma}}_n^{-1}\mathbf{1}_p}.
\end{equation}

\citet{okhrin2006distributional} derive the analytical expression for the expectation  and the covariance matrix of $\hat{\mathbf{w}}_{EU}$ and obtain its asymptotic distribution assuming that the portfolio size is considerably smaller than the sample size. These results are extended in \citet{bodnarEU} who derive the finite-sample distribution of the estimated EU portfolio weights and use these results in the derivation of an asymptotic tests on the weights which we present in the next subsection.

\subsection{Tests based on Mahalanobis distance}\label{sec2a}
At each time point an investor has to decide whether the holding portfolio is efficient or it has to be adjusted (see, \citet{bodnar2008test}, \citet{bodnarEU}). This problem can be presented as a special case of the general linear hypotheses formulated for the portfolio weights. Let $\mathbf{L}$ denote the $k\times p$ dimensional matrix of constants with $k<p-1$ and let $\mathbf{r}$ be the $k$-dimensional vector of constants. \citet{bodnarEU} consider the following hypotheses for linear combinations of the EU portfolio weights
\begin{equation}\label{hypotheses}
H_{0}:\mathbf{L}\mathbf{w}_{EU}=\mathbf{r}\qquad \textrm{against} \qquad H_{1}:\mathbf{L}\mathbf{w}_{EU}\not= \mathbf{r},
\end{equation}
If one sets $\mathbf{L}=[\mathbf{I}_{k} ~ \mathbf{O}_{k,p-k}]$ in \eqref{hypotheses} where $\mathbf{I}_{k}$ is the $k$-dimensional identity matrix and $\mathbf{O}_{k,p-k}$ is the $k \times (p-k)$ matrix with zeros, then the null hypothesis states that the first $k$ weights in $\mathbf{w}_{EU}$ are equal to the corresponding components defined by $\mathbf{r}$. It also has to be noted that whole structure of the EU portfolio cannot be tested by using \eqref{hypotheses} because of the restriction imposed on the number of linear combinations which should be smaller than $p-1$. Thus, the test on the whole vector of the EU portfolio weights should be performed by testing at least two null hypotheses of the form \eqref{hypotheses} by selecting matrices $\mathbf{L}$ in each of the null hypotheses such that all elements in $\mathbf{w}_{EU}$ are tested. This leads to a multiple testing problem also discussed below.

In order to test \eqref{hypotheses} for a given matrix $\mathbf{L}$ and a vector $\mathbf{r}$, \citet{bodnarEU} suggest the following test statistic:
{\footnotesize\begin{eqnarray}\label{Stat1}
 T_{\mathbf{L}}&=& (n-p+1)\left(\hat{\mathbf{w}}_{\mathbf{L}}- \mathbf{r} \right)'\left( \frac{\mathbf{L}\hat{\mathbf{Q}}_n\mathbf{L}'}{\textbf{1}_p'\hat{\mathbf{\Sigma}}_n^{-1}\mathbf{1}_p} + \gamma^{-1}\frac{\mathbf{L}\hat{\mathbf{Q}}_n\mathbf{L}'}{\mathbf{\bar{x}}_{n}'\hat{\mathbf{Q}}_n\mathbf{\bar{x}}_{n}}
+\gamma^{-2}(\mathbf{L}\hat{\mathbf{Q}}_n\mathbf{L}'\mathbf{\bar{x}}_{n}'\hat{\mathbf{Q}}_n\mathbf{\bar{x}}_{n} - \mathbf{L}\hat{\mathbf{Q}}_n \mathbf{\bar{x}}_{n}\mathbf{\bar{x}}_{n}'\hat{\mathbf{Q}}_n\mathbf{L}')\right)^{-1}\nonumber\\
&\times& \left(\hat{\mathbf{w}}_{\mathbf{L}}- \mathbf{r} \right),
\end{eqnarray}}
where
\begin{equation}\label{hwL}
\hat{\mathbf{w}}_{\mathbf{L}} = \mathbf{L} \hat{\mathbf{w}}_{EU}=  \frac{\mathbf{L}\hat{\mathbf{\Sigma}}_n^{-1}\textbf{1}_p}{\textbf{1}_p'\hat{\mathbf{\Sigma}}_n^{-1}\mathbf{1}_p}+ \gamma^{-1}\mathbf{L}\hat{\mathbf{Q}}_n\mathbf{\bar{x}}_{n}.
\end{equation}

\citet{bodnarEU} show that the test statistic $T_{\mathbf{L}}$ can be asymptotically well approximated by a non-central $\chi^2$-distribution with $k$ degrees of freedom and the non-centrality parameter

\begin{equation}
\lambda= n \left(\mathbf{w}_{\mathbf{L}}- \mathbf{r} \right)'\left( \frac{\mathbf{L}\mathbf{Q}\mathbf{L}'}{\textbf{1}_p'\mathbf{\Sigma}^{-1}\mathbf{1}_p} +\gamma^{-1}\frac{\mathbf{L}\mathbf{Q}_n\mathbf{L}'}{\boldsymbol{\mu}'\mathbf{Q}\boldsymbol{\mu}}
+\gamma^{-2}(\mathbf{L}\mathbf{Q}\mathbf{L}'\boldsymbol{\mu}'\mathbf{Q}\boldsymbol{\mu} - \mathbf{L}\mathbf{Q} \boldsymbol{\mu}\boldsymbol{\mu}'\mathbf{Q}\mathbf{L}')\right)^{-1} \left(\mathbf{w}_{\mathbf{L}}- \mathbf{r} \right)
\end{equation}
with
\begin{equation}\label{wL}
\mathbf{w}_{L} = \mathbf{L} \mathbf{w}_{EU}=  \frac{\mathbf{L}\mathbf{\Sigma}^{-1}\textbf{1}_p}{\textbf{1}_p'\mathbf{\Sigma}^{-1}\mathbf{1}_p}+ \gamma^{-1}\mathbf{L}\mathbf{Q}\boldsymbol{\mu},
\end{equation}
when both $p$ and $k$ are relatively small with respect to the sample size $n$. As a special case, we obtain the asymptotic distribution of $T_{\mathbf{L}}$ under the null hypothesis. This appears to be a $\chi^2$-distribution, i.e. $T_{\mathbf{L}}\sim \chi^2_k$ under the null hypothesis in \eqref{hypotheses}.

Since the asymptotic distribution of the test statistic $T_{\mathbf{L}}$ is obtained under classical asymptotic regime, this test, in general, is not applicable when the portfolio size is comparable to the sample size. We illustrate this point in Figure \ref{fig1}. Here we plot the kernel density estimator (KDE) of the distribution of the test statistic $T_{\mathbf{L}}$ under the null hypothesis together with the asymptotic $\chi^2$-distribution (green and red curves, respectively). For this purpose we generate samples from a multivariate normal distribution with mean vector and covariance matrix as specified in the numerical study of Section~\ref{sec4a}. The vector $\mathbf{r}$ consists of the first $k$ components of the true EU portfolio weights and we set $\mathbf{L}=[\mathbf{I}_{k} ~ \mathbf{O}_{k,p-k}]$. For each sample we compute  the value of the test statistic $T_{\mathbf{L}}$ and then plot the KDE. To robustify the conclusions we set $\gamma =5$, $p=300$, $c_n=p/n \in\{0.3, 0.8\}$ and  $k \in \{10,  30, 100\}$. We observe that already for $k=10$ the difference between the KDE and the asymptotic distribution is very large and this evidence becomes stronger if $k$ increases. For $k=100$ the KDE shifts strongly to the right and is not shown to retain the same scaling on the $x$-axis. Table \ref{table: tab1} gives the realized sizes (type I errors) of the considered test based on the $5000$ independent replications and with the nominal level $\alpha=0.05$. For different values of $k \in \{10,  30, 100\}$, it can be seen that  $T_{\mathbf{L}}$ is highly inconsistent and has a much higher size than the nominal value $\alpha$.  We conclude that the test is highly unreliable if we wish to test many or all weights simultaneously.

\subsection{Improvement of the test based on Mahalanobis distance for large-dimensional portfolios}\label{sec2b}

\citet{bodnar2019sampling} show that the sample estimator of the EU portfolio weights is not consistent under the high-dimensional asymptotic regime, i.e., when $p/n \to c \in[0,1)$ as $n \to  \infty$. Moreover, they derive a consistent estimator for the elements of $\mathbf{w}_{EU}$ and use these findings to construct a high-dimensional asymptotic test on the finite number of linear combinations of the EU portfolio weights.

Let $\mathbf{L}$ be a $k\times p$ matrix of constant as defined in Section~\ref{sec2a} and let
\begin{equation*}\label{hwLs-eta}
\hat{\mathbf{w}}_{GMV;\mathbf{L}} = \mathbf{L} \hat{\mathbf{w}}_{GMV}=  \frac{\mathbf{L}\hat{\mathbf{\Sigma}}_n^{-1}\textbf{1}_p}{\textbf{1}_p'\hat{\mathbf{\Sigma}}_n^{-1}\mathbf{1}_p}, \quad
\hat{s}=\mathbf{\bar{x}}_{n}'\hat{\mathbf{Q}}_n\mathbf{\bar{x}}_{n}
\end{equation*}
\begin{equation}
\mbox{and} \quad
\hat{\boldsymbol{\eta}}_{\mathbf{L}} =\frac{\mathbf{L}\hat{\mathbf{Q}}_n\mathbf{\bar{x}}_{n}}{\mathbf{\bar{x}}_{n}'\hat{\mathbf{Q}}_n\mathbf{\bar{x}}_{n}}.
\end{equation}
Assuming that $k$ is finite, i.e. considerably smaller than both $p$ and $n$, \citet{bodnar2019sampling} prove that
\begin{equation}\label{hwLs-eta-as}
\hat{\mathbf{w}}_{GMV;\mathbf{L}} \stackrel{a.s.}{\to} \mathbf{L} \mathbf{w}_{GMV}, \quad
\hat{s}_c=(1-c_n)\hat{s}-c_n \stackrel{a.s.}{\to} s
\end{equation}
\begin{equation}
\mbox{and} \quad
\hat{\boldsymbol{\eta}}_{\mathbf{L};c}= \frac{\hat{s}_c+c_n}{\hat{s}_c} \hat{\boldsymbol{\eta}}_{\mathbf{L}} \stackrel{a.s.}{\to}  \boldsymbol{\eta}_{\mathbf{L}}
\end{equation}
for $c_n=p/n \to c \in[0,1)$ as $n \to \infty$ with
\begin{equation}\label{s-eta}
s= \boldsymbol{\mu}'\mathbf{Q} \boldsymbol{\mu}
\quad \mbox{and} \quad
\boldsymbol{\eta}_{\mathbf{L}}= \frac{\mathbf{L}\mathbf{Q} \boldsymbol{\mu}}{\boldsymbol{\mu}'\mathbf{Q} \boldsymbol{\mu}}.
\end{equation}
The symbol $\stackrel{a.s.}{\to}$ denotes the almost surely convergence.

Using \eqref{hwLs-eta-as}, \citet{bodnar2019sampling} propose a high-dimensional asymptotic test on the hypotheses \eqref{hypotheses} with the test statistic given by
\begin{equation}\label{Stat1c}
T_{\mathbf{L};c}= (n-p)\left(\hat{\mathbf{w}}_{\mathbf{L};c}- \mathbf{r} \right)'\hat{\mathbf{\Omega}}_{\mathbf{L};c}^{-1}
\left(\hat{\mathbf{w}}_{\mathbf{L};c}- \mathbf{r} \right),
\end{equation}
where
\begin{equation}\label{hwLc}
\hat{\mathbf{w}}_{\mathbf{L};c} = \hat{\mathbf{w}}_{GMV;\mathbf{L}}+ \gamma^{-1}\hat{s}_c \hat{\boldsymbol{\eta}}_{\mathbf{L};c}
\end{equation}
and
\begin{equation}
	\begin{split}\label{hOmegac}
	\hat{\mathbf{\Omega}}_{\mathbf{L};c} &=
	\bigg(
	\left(\frac{1-c_n}{\hat{s}_c+c_n} + (\hat{s}_c+c_n)\gamma^{-1} \right)\gamma^{-1} + \hat{V}_{c}\bigg) (1-c_n)\mathbf{L} \hat{\mathbf{Q}}_n \mathbf{L}^\top\\
	&+ {\tiny \gamma^{-2}\Bigg\{\frac{2(1-c_n)c_n^3 }{(\hat{s}_c+c_n)^2}+  4(1-c_n)c_n\frac{\hat{s}_c(\hat{s}_c+2c_n)}{(\hat{s}_c+c_n)^2} 
	+  \frac{2(1-c_n)c_n^2(\hat{s}_c+c_n)^2}{\hat{s}_c^2} -\hat{s}_c^2  \Bigg\} \hat{\boldsymbol{\eta}}_{\mathbf{L};c}\hat{\boldsymbol{\eta}}_{\mathbf{L};c}'},
	\end{split}
	\end{equation}
	
	where
	\begin{equation}\label{hVc}
	\hat{V}_{c}=\frac{\hat{V}_{GMV}}{1-c_n} \quad \mbox{with} \quad
	\hat{V}_{GMV}=\frac{1}{\textbf{1}_p'\hat{\mathbf{\Sigma}}_n^{-1}\mathbf{1}_p}
	\end{equation}
are the consistent and the sample estimators of the variance of the GMV portfolio \eqref{solution_GMV}, that is (see, e.g., \citet[p.387]{bodnar2018estimation})
\[\hat{V}_{c}\stackrel{a.s.}{\to}V_{GMV}=\frac{1}{\textbf{1}_p'\mathbf{\Sigma}^{-1}\mathbf{1}_p}\]
for $c_n=p/n \to c \in[0,1)$ as $n \to \infty$.

The application of the results of Theorem 4.4 in \citet{bodnar2019sampling} leads to the high-dimensional asymptotic distribution of $T_{\mathbf{L};c}$ under both hypotheses in \eqref{hypotheses}. Namely, it holds that the asymptotic distribution of $T_{\mathbf{L};c}$ under $H_1$ is well approximated by a non-central $\chi^2$-distribution with $k$ degrees of freedom and non-centrality parameter given by
\begin{equation}
\lambda_c=(n-p)(\mathbf{w}_{\mathbf{L}}- \mathbf{r})'\mathbf{\Omega}_{\mathbf{L};c}^{-1}(\mathbf{w}_{\mathbf{L}}- \mathbf{r}),
\end{equation}
where
	\begin{eqnarray}\label{Omegac}
	\mathbf{\Omega}_{\mathbf{L};c} &=&
	\bigg(
	\left(\frac{1-c}{s+c} + (s+c)\gamma^{-1} \right)\gamma^{-1} + V_{GMV} \bigg) (1-c)\mathbf{L} \mathbf{Q} \mathbf{L}^\top\nonumber \\
	&+& \gamma^{-2}\Bigg\{\frac{2(1-c)c^3 }{(s+c)^2}+  4(1-c)c\frac{s(s+2c)}{(s+c)^2} \nonumber+  \frac{2(1-c)c^2(s+c)^2}{s^2} -s^2  \Bigg\} \boldsymbol{\eta}_{\mathbf{L}}\boldsymbol{\eta}_{\mathbf{L}}'.
	\end{eqnarray}

Moreover, $T_{\mathbf{L};c} \stackrel{d}{\to} \chi^2_k$ under $H_0$, where the symbol $\stackrel{d}{\to}$ denotes the convergence in distribution.
\begin{figure}[!h]
	\centering
	
	\begin{tabular}{cc}	\hspace{-0.2cm}	\includegraphics[scale=0.45]{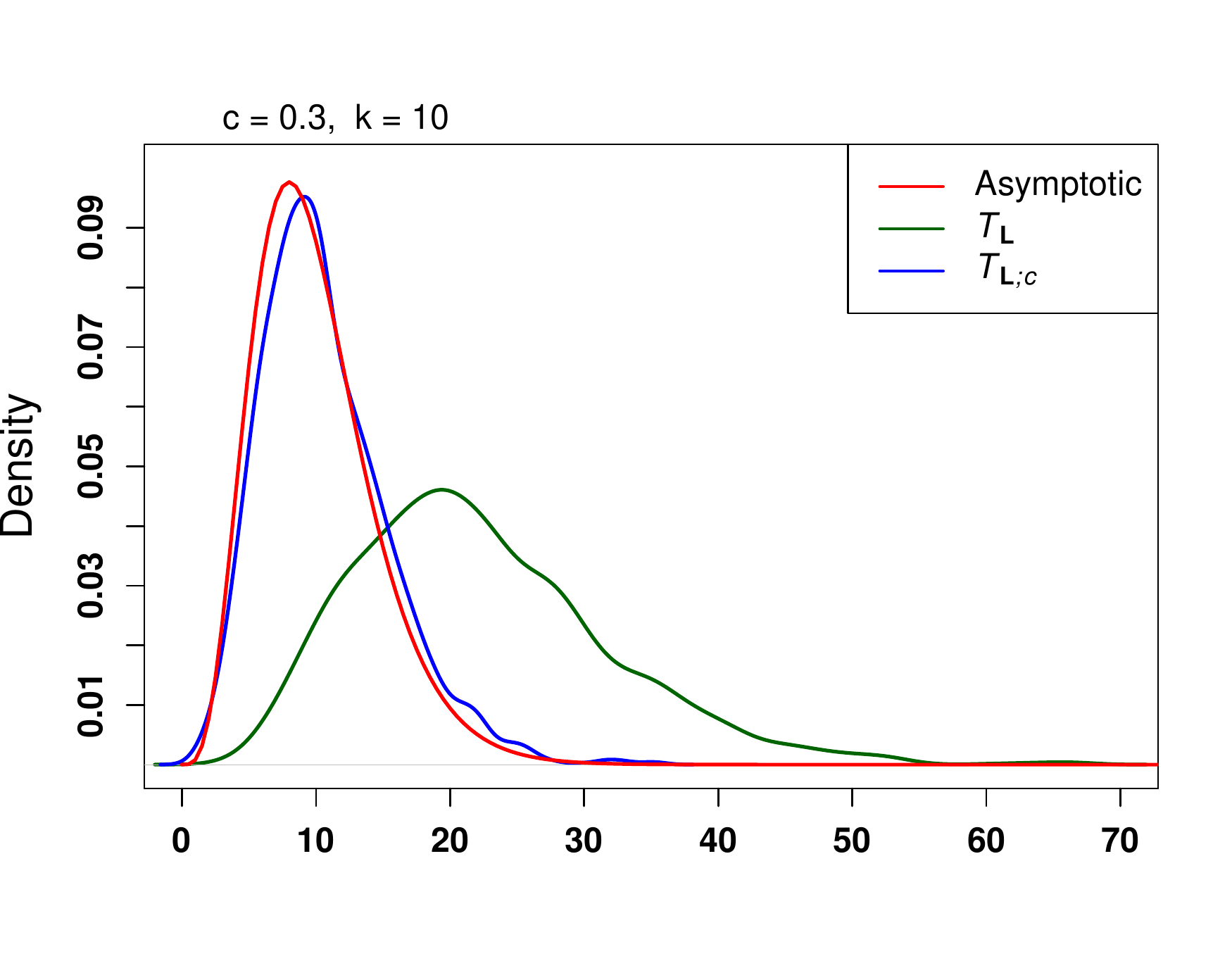}& 	
		\includegraphics[scale=0.45]{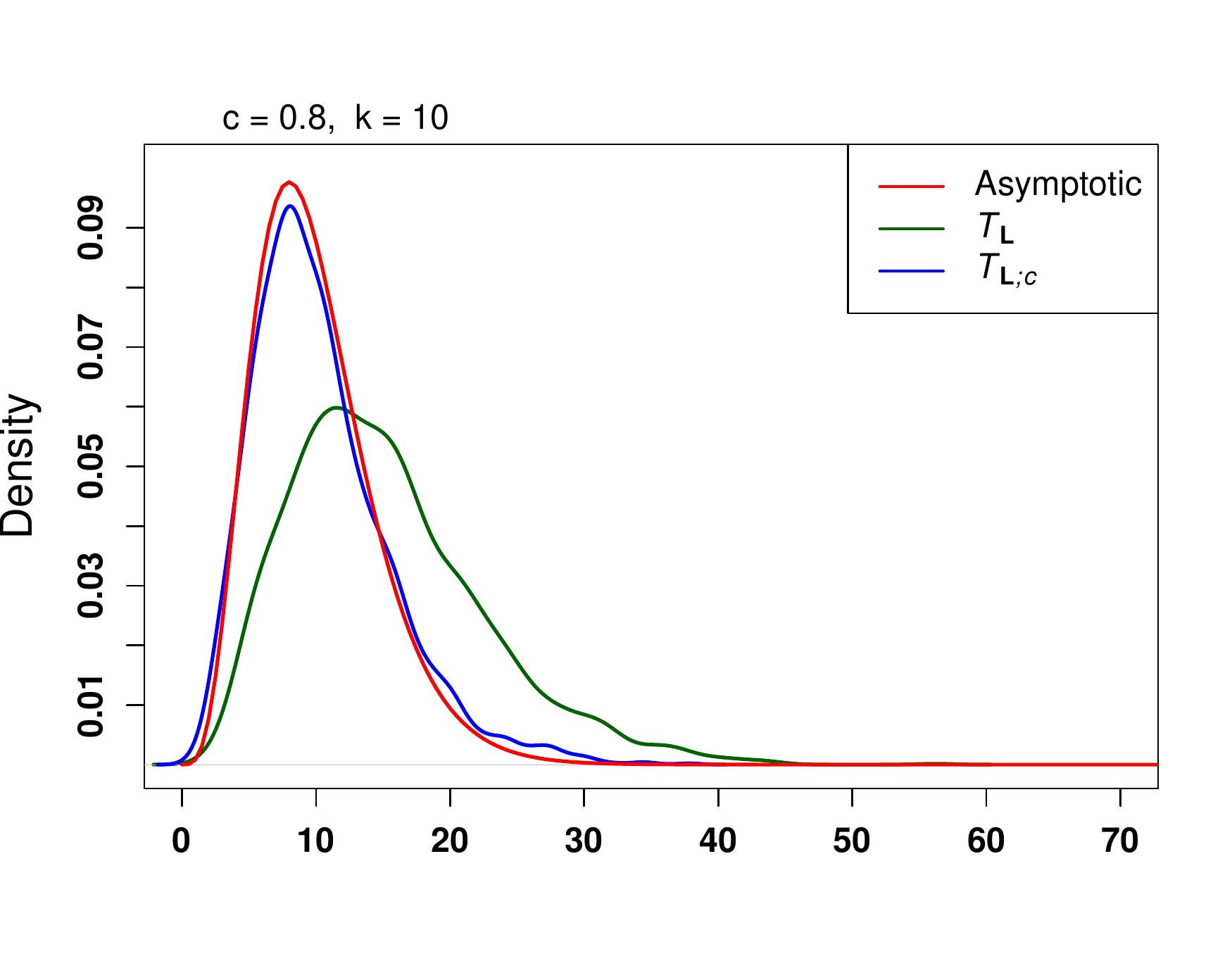}\\
		\includegraphics[scale=0.45]{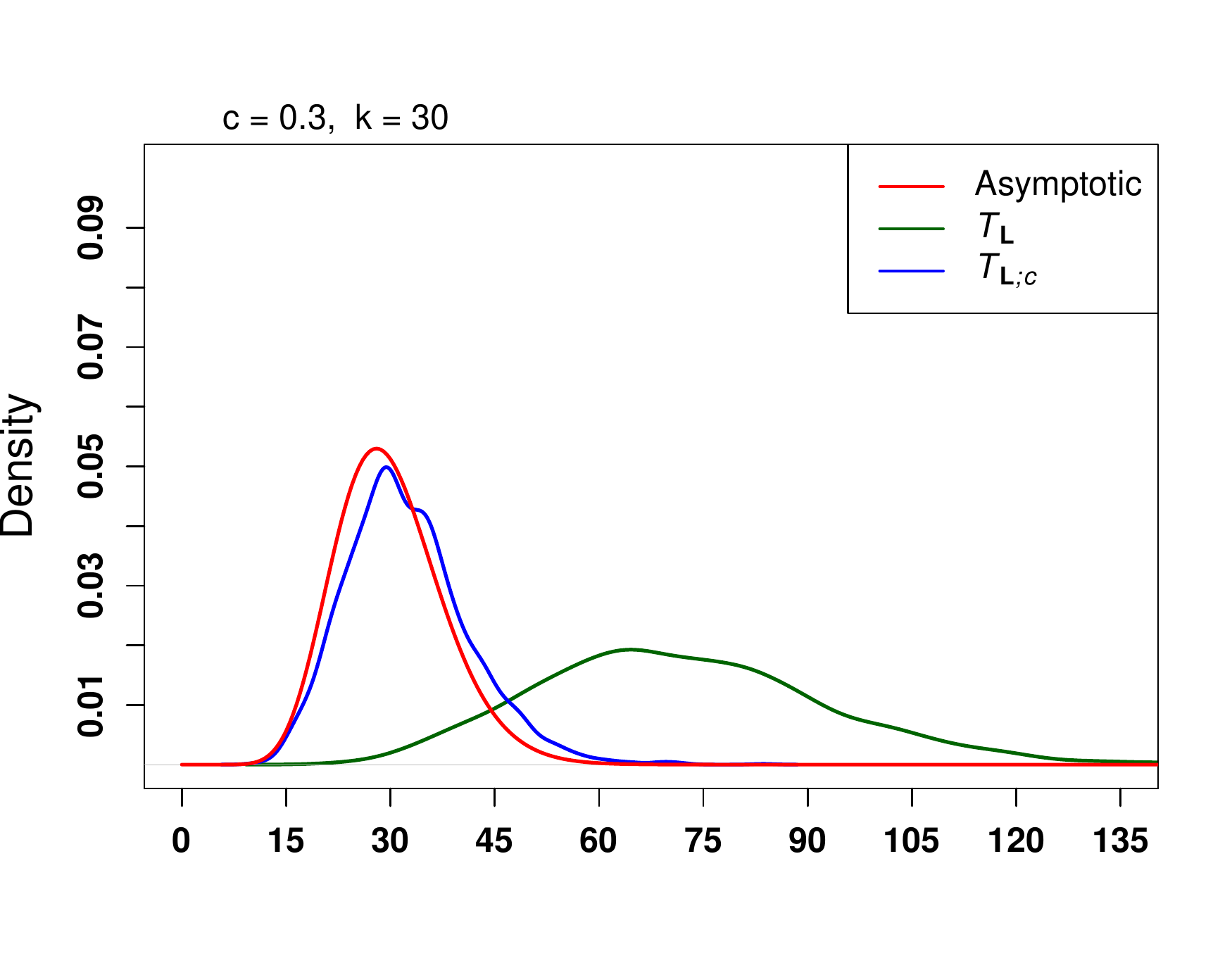}&  	
		\includegraphics[scale=0.45]{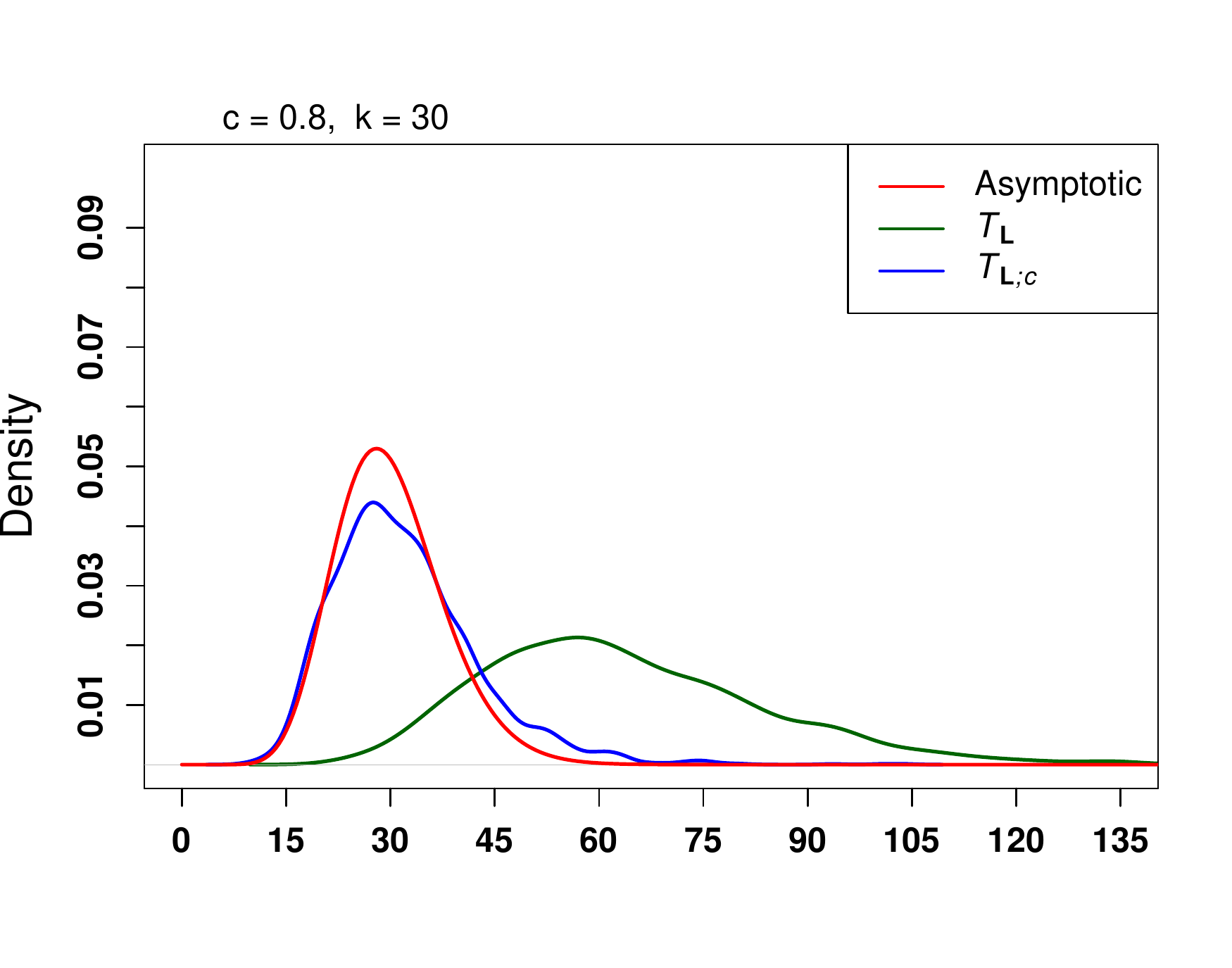}\\
		\includegraphics[scale=0.45]{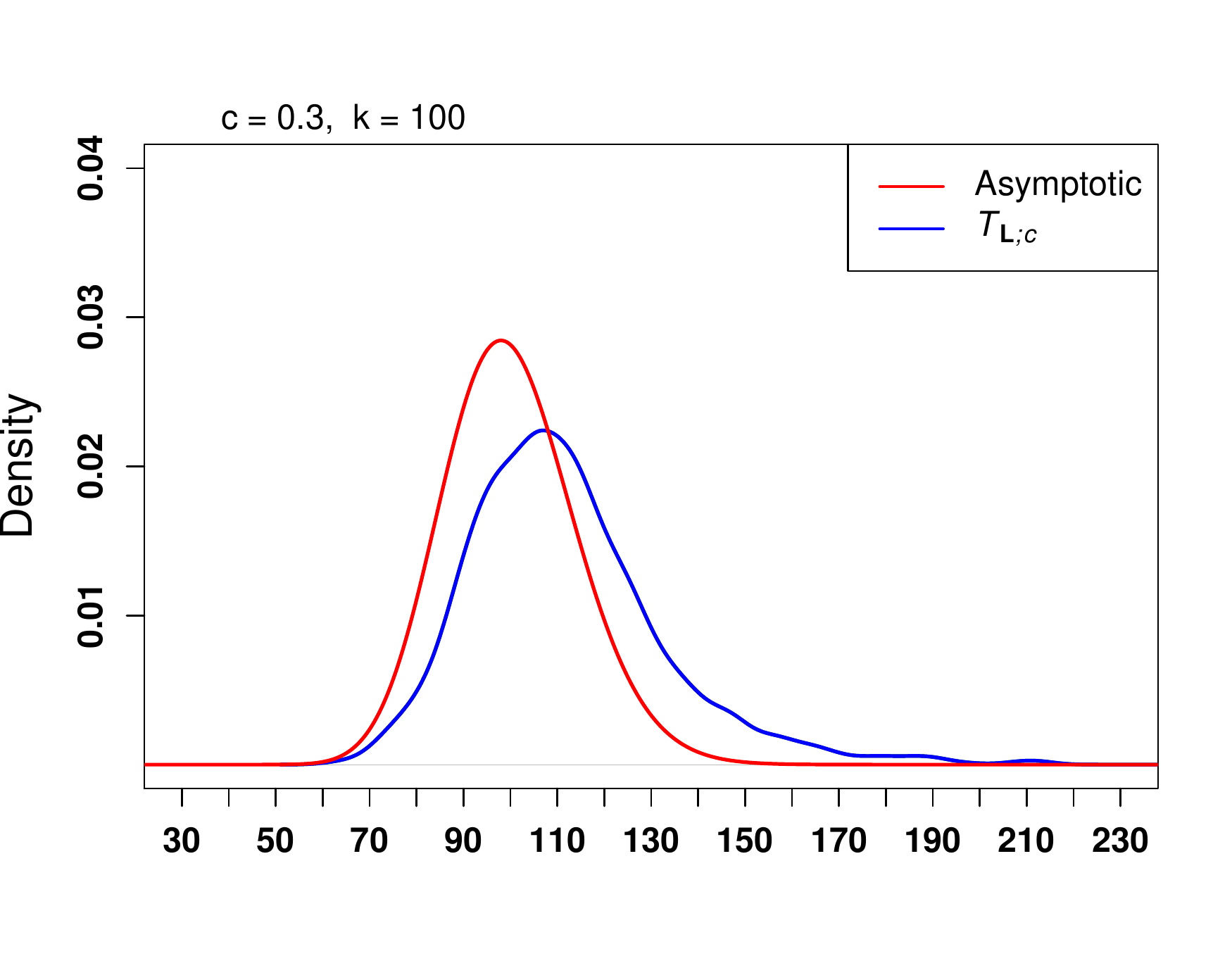}&  	
		\includegraphics[scale=0.45]{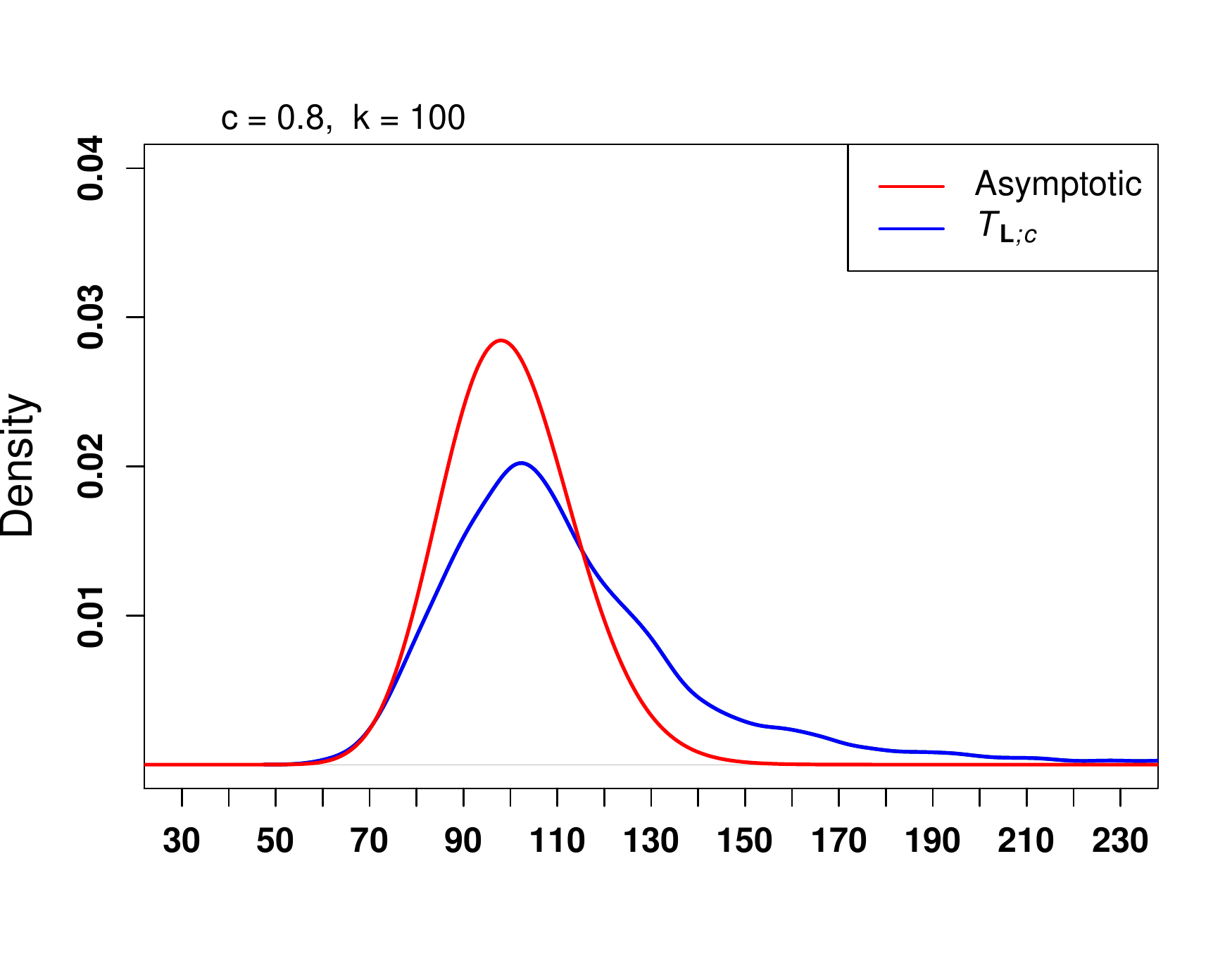}\\
	\end{tabular}
	
	\caption{The high-dimensional asymptotic $\chi^2$ approximation of the densities of $T_{\mathbf{L}}$  and $T_{\mathbf{L};c}$ together with their kernel density estimators for $\gamma =5$, $p=300$,  $c_n=p/n \in\{0.3,0.8\}$ and $k \in \{10, 30, 100\}$.  } \label{fig1}
\end{figure}

In Figure $\ref{fig1}$ we present the KDE of the distribution of $T_{\mathbf{L};c}$ (blue curve) and compare it to its high-dimensional asymptotic distribution (red curve). The kernel density estimator as well as the sizes of the test are obtained under the same simulation setup as one used at the end of Section 2.1. The approximation works well and much better than in the case of $T_{\mathbf{L}}$ for smaller values of $k$, but discrepancy becomes large if $k$ increases.  The same conclusion can be drawn from Table \ref{table: tab1}. Here  the method proposed by \citet{bodnar2019sampling} has a much better realized size which still increases  dramatically with growing $k$.

\begin{table}[!h]
	\renewcommand{\arraystretch}{1.3}
	\centering	
	\begin{tabular}{lccc}	
		\multicolumn{4}{c}{$\mathbf{c=0.3}$}                                                  \\ \hline \hline
		& \multicolumn{1}{l}{$k=10$} & \multicolumn{1}{l}{$k=30$} & \multicolumn{1}{l}{$k=100$} \\ \cline{2-4}
		$T_{\mathbf{L}}$	& 0.528      & 0.891       & 1             \\ \cline{2-4}
		$T_{\mathbf{L};c}$	& 0.061      & 0.071       & 0.181       \\ \hline
		\multicolumn{4}{c}{$\mathbf{c=0.8}$}   \\ \hline \hline
		& $k=10$                    & $k=30 $  & $k=100$     \\ \cline{2-4}
		$T_{\mathbf{L}}$	    & 0.226         & 0.765    & 1           \\ \cline{2-4}
		$T_{\mathbf{L};c}$		& 0.069         & 0.105    & 0.221        \\ \hline
	\end{tabular} 
	\caption{Empirical sizes of the tests based on $T_{\mathbf{L}}$ and $T_{\mathbf{L};c}$ using $5 \cdot 10^3 $ independent replications.}  \label{table: tab1}
\end{table}
\section{Test based on the shrinkage approach}\label{sec3}

Both tests based on the Mahalanobis distance are designed to test a finite number of linear restrictions imposed on the EU portfolio weights. Although the high-dimensional test shows a considerable improvement in terms of the size (see, Figures \ref{fig1} and Table \ref{table: tab1}), this test, similarly to the test based on the statistic $T_{\mathbf L}$, cannot be applied to test the structure of the whole EU portfolio. In practice, one has to fix the number $k$ of the EU portfolio weights (or their linear restrictions) and apply the test $T_{\mathbf{L};c}$ several times in order to cover the whole vector $\mathbf{w}_{EU}$. This approach is a single-step multiple test (see, \citet{dickhaus2014simultaneous}) with the number of marginal hypotheses to be tested equal to $[p/k]+1$. Since the dependence structure between the marginal tests is very complicated, one has to monitor the overall type I error rate by using the so-called Bonferroni correction (see, \citet{dickhaus2014simultaneous}). This would worse the power properties of each individual test, especially when the number of tests is relatively large.

As a solution to this challenging problem, we suggest a new approach for testing the structure of the EU portfolio by a single test. The new procedure is based on the shrinkage estimator of the EU portfolio weights as suggested by \citet{bodnar2016okhrin} and extend our previous results obtained for the GMV portfolio in \citet{BodnarDmytriv2019}, which is a very special case of the EU portfolio. In contrast to the EU portfolio, the weights of the GMV portfolio do not depend on the mean vector. As a result, the derivation of the test for the EU portfolio becomes a very challenging task and completely new results in random matrix theory have to be derived to handle it.

\subsection{Optimal shrinkage estimator of the EU portfolio weights}\label{sec3a}

The shrinkage estimator for the EU portfolio weights is a convex combination of the sample estimator and a fixed well behaved target portfolio $\mathbf{b} \in \mathbb{R}^p$ with bounded expected return and variance, i.e., $R_b=\mathbf{b}'\boldsymbol{\mu}<\infty$ and $V_b=\mathbf{b}'\boldsymbol{\Sigma}^{-1}\mathbf{b}<\infty$ uniformly in $p$. Thus, the shrinkage estimator is expressed as
\begin{equation}\label{hwGSE}
\hat{\mathbf{w}}_{GSE}= \alpha_n \hat{\mathbf{w}}_{EU} + (1-\alpha_n)\mathbf{b} \quad \mbox{with} \quad \mathbf{b}'\mathbf{1}_p=1,
\end{equation}
where $\alpha_n$ is the shrinkage intensity. One of the main ideas behind the shrinkage estimator \eqref{hwGSE} is to reduce the large variability present in the sample estimator $\hat{\mathbf{w}}_{EU}$ by shrinking it to a vector of constants. This approach might introduce a bias in the estimator, but on the other side it reduces the variability of the sample estimator considerably.

\citet{bodnar2016okhrin} determine the optimal shrinkage intensity $\alpha^{*}_n$ as the solution of the maximization problem based on the mean-variance objective function. It is given by
\begin{equation}\label{opt_shrink}
\alpha^{*}_n=\frac{(\hat{\mathbf{w}}_{EU}- \mathbf{b})'(\boldsymbol{\mu}-\gamma\mathbf{\Sigma}\mathbf{b})}{(\hat{\mathbf{w}}_{EU}- \mathbf{b})'\mathbf{\Sigma}(\hat{\mathbf{w}}_{EU}- \mathbf{b})}
\end{equation}
Since the expression of $\alpha^{*}_n$ depends on both the population mean vector and covariance matrix and on their sample counterparts, it cannot be directly applied in practice. As such, \citet{bodnar2016okhrin} propose a two-stage procedure. First, the deterministic quantity $\alpha^{*}$ which is asymptotically equivalent to $\alpha^{*}_n$ is found. Second, it is consistently estimated under the high-dimensional asymptotic regime.

It holds that (see, \citet[Theorem 2.1]{bodnar2016okhrin})
\begin{equation}\label{alpha_opt}
\alpha^{*} =\gamma^{-1} \frac{(R_{GMV}-R_b)\left( 1+ \frac{1}{1-c}\right)+\gamma(V_b-V_{GMV})+\frac{\gamma^{-1}}{1-c}s }{\frac{1}{1-c}V_{GMV} -2\left(V_{GMV}+\frac{\gamma^{-1}}{1-c}(R_b- R_{GMV}) \right) + \gamma^{-2}\left(\frac{s}{(1-c)^3}+\frac{c}{(1-c)^3} \right)+ V_b },
\end{equation}
where $R_{GMV}=\frac{\mathbf{1}_p'\mathbf{\Sigma}^{-1}\boldsymbol{\mu}}{\mathbf{1}_p'\mathbf{\Sigma}^{-1}\mathbf{1}_p}$ is the expected return of the GMV portfolio.
Following \citet{bodnar2016okhrin} we assume throughout the paper that uniformly in $p$ the quadratic form $\mathbf{1}'\mathbf{\Sigma}^{-1}\mathbf{1}_p$ is bounded away from zero and $\boldsymbol{\mu}'\mathbf{\Sigma}^{-1}\boldsymbol{\mu}$ is bounded from above by some positive constant. These conditions guarantee among others the boundedness of $R_{GMV}$, $V_{GMV}$ and $s$ as $p\to\infty$, thus, keeping the limiting expressions coming further well defined asymptotically.
Consistent estimators for the variance of the GMV portfolio $V_{GMV}$ and for the slope parameter of the efficient frontier $s$ are given in \eqref{hVc} and \eqref{hwLs-eta-as}, respectively. \citet{bodnar2016okhrin} show that the sample estimators of $R_{GMV}$, $R_b$, and $V_b$ are consistent, that is
\begin{equation}\label{consist_est}\begin{split}
\hat{R}_{GMV}&=\dfrac{\mathbf{1}_p'\hat{\mathbf{\Sigma}}_n^{-1}\mathbf{\bar{x}}_{n}}{\mathbf{1}_p'\hat{\mathbf{\Sigma}}_n^{-1}\mathbf{1}_p} \stackrel{a.s}{\to} R_{GMV},\\
&\hspace{-0.4cm} \hat{R_b}= \mathbf{b}'\mathbf{\bar{x}}_{n} \stackrel{a.s}{\to} R_b,\\
&\hspace{-0.31cm}\hat{V_b}=\mathbf{b}'\hat{\mathbf{\Sigma}}_n\mathbf{b}\stackrel{a.s}{\to} V_b,
\end{split}
\end{equation}
for $p/n \to c\in[0,1)$ as $n \to  \infty$.

Hence, a consistent estimator for $\alpha^{*} $ is constructed as
 \begin{equation}\label{alpha_est}
 \hat{\alpha}_c^* = \gamma^{-1}\frac{(\hat{R}_{GMV}-\hat{R}_b) \left( 1+\frac{1}{1-c_n}\right)+\gamma(\hat{V}_b-\hat{V}_c) + \frac{\gamma^{-1}}{1-c_n}\hat{s}_c} {\frac{1}{1-c_n}\hat{V}_c-2\left(\hat{V}_c+\frac{\gamma^{-1}}{1-c_n}(\hat{R}_b-\hat{R}_{GMV}) \right)+ \gamma^{-2}\left(\frac{\hat{s}_c}{(1-c_n)^3} +\frac{c_n}{(1-c_n)^3}\right)+ \hat{V}_b},
 \end{equation}
while the bona fide shrinkage estimator for the weights of the EU portfolio are expressed as
\begin{equation}
\hat{\mathbf{w}}_{BFGSE}= \hat{\alpha}_c^* \hat{\mathbf{w}}_{EU} + (1- \hat{\alpha}_c^*)\mathbf{b}.
\end{equation}
Next, we prove that $ \hat{\alpha}_c^*$ is asymptotically normally distributed. This result will then be used to derive a test for the structure of the EU portfolio in Section\ref{sec3b}.
Let $\alpha^*=\frac{A}{B}$ and $\hat{\alpha}_c^*=\frac{\hat{A}_n}{\hat{B}_n}$. Then, we get
\begin{eqnarray}\label{dif_alp}
&&\sqrt{n}(\hat{\alpha}_c^*-\alpha^*)
= \sqrt{n}\left(\frac{\hat{A}_n-A}{\hat{B}_n}-\frac{A(\hat{B}_n-B)}{B\hat{B}_n}\right)\nonumber\\
&=& \frac{1}{\hat{B}_n}\left(\sqrt{n}(\hat{A}_n-A)-\frac{A}{B}\sqrt{n}(\hat{B}_n-B)\right)\nonumber\\
&=& \frac{\mathbf{d}'}{\hat{B}_n} \sqrt{n} \mathbf{t} + o_P(1)
\end{eqnarray}
for $p/n \to c+o(n^{-1/2})$ as $n \to \infty$ with

	\begin{equation}\label{bt}
	\mathbf{t}=
	\begin{pmatrix}
	\hat{R}_{GMV}- R_{GMV}\\
	\hat{V}_{c}-V_{GMV}\\
	\hat{s}_c-s\\
	\hat{R}_b-R_b\\
	\hat{V}_{b}-V_b
	\end{pmatrix}
	~\text{and}~
	\mathbf{d}=
	\begin{pmatrix}
	1+\frac{1}{1-c_n}\left(1-2\frac{A}{B}\right)\\
	-\gamma \left(1 +\frac{A}{B}\left(\frac{1}{1-c_n}-2\right)\right)\\
	\frac{\gamma^{-1}}{1-c_n}\left(1-\frac{1}{(1-c_n)^2}\frac{A}{B}\right)\\
	-1-\frac{1}{1-c_n}\left(1-2\frac{A}{B}\right)\\
	\gamma \left(1 -\frac{A}{B}\right)
	\end{pmatrix},
	\end{equation}
where the symbol $o_P(1)$ denotes a sequence which tends almost surely to zero. In Theorem \ref{th1} we derive the asymptotic distribution of $\mathbf{t}$.

\begin{theorem}\label{th1}
	Let $\mathbf{x}_1, \ldots, \mathbf{x}_n$ be independent and identically distributed with $\mathbf{x}_i\sim \mathcal{N}_p(\boldsymbol{\mu}, \mathbf{\Sigma})$ for $i=1,\ldots,n$ with $\mathbf{\Sigma}$ positive definite. Then it holds that
	\begin{equation}\label{asym_bt}
	\sqrt{n}\mathbf{t} \stackrel{d}{\to} \mathcal{N}_5(\mathbf{0},\mathbf{\Omega}_{\alpha})
	\end{equation}
	for $p/n \to c\in [0,1)$ as $n \to \infty$ where
	
	%\vspace{1.5cm}

		\begin{equation}\label{asym_bt_var}
		\mathbf{\Omega}_{\alpha}=\begin{pmatrix}
		\frac{V_{GMV}(s+1)}{1-c} & 0 &0&V_{GMV}&-2V_{GMV}(R_b-R_{GMV})\\
		0 & 2\frac{V_{GMV}^2}{1-c}& 0&0&2V_{GMV}^2\\
		0& 0&2\frac{((s+1)^2+c-1)}{1-c}&2(R_b-R_{GMV})&-2(R_b-R_{GMV})^2\\
		V_{GMV}&0&2(R_b-R_{GMV})&V_b&0\\
		-2V_{GMV}(R_b-R_{GMV})&2V_{GMV}^2&-2(R_b-R_{GMV})^2&0&2V_b^2\\
		\end{pmatrix}.
		\end{equation}

\end{theorem}

\vspace{-0.6cm}Since
\[\hat{B}_n \stackrel{a.s}{\to}B
\quad \mbox{for} \quad \frac{p}{n} \to c\in[0,1) \quad \mbox{as} \quad n \to \infty, \]
the application of Slutsky's lemma (c.f., \citet[Theorem 1.5]{dasgupta2008}) leads to the asymptotic distribution of $\hat\alpha_c^*$ as given in Theorem \ref{th2}.

\begin{theorem}\label{th2}
	Under the assumptions of Theorem \ref{th1}, it holds that
	\begin{equation}\label{asym_alpha}
	\sqrt{n}(\hat{\alpha}_c^*-\alpha^*) \stackrel{d}{\to} \mathcal{N}(0,C_{\alpha}),
	\end{equation}
	for $p/n \to c\in [0,1)$ as $n \to \infty$ where
	\begin{equation}\label{asym_sig_alp}
	C_{\alpha}=\frac{1}{B^2}\mathbf{d}'\mathbf{\Omega}_{\alpha} \mathbf{d}\,.
	\end{equation}
\end{theorem}

Finally, using \eqref{hwLs-eta-as}, \eqref{hVc}, and \eqref{consist_est} a consistent estimator for $C_{\alpha}$ is given by
\begin{equation}\label{asym_sig_alp_c}
\hat C_{\alpha}=\frac{1}{\hat B_n^2}\mathbf{d}'\hat{\mathbf{\Omega}}_{\alpha;c} \mathbf{d}\,,
\end{equation}
where $\hat{\mathbf{\Omega}}_{\alpha;c}$ is a consistent estimator for $\mathbf{\Omega}_{\alpha}$ expressed as

	\begin{equation}\label{asym_sig_alp_c2}
	\hat{\mathbf{\Omega}}_{\alpha;c}=
	\begin{pmatrix}
	\frac{\hat{V}_{c}(\hat{s}_c+1)}{1-c} & 0 &0&\hat{V}_{c}&-2\hat{V}_{c}(\hat{R}_b-\hat{R}_{GMV})\\
	0 & 2\frac{\hat{V}_{c}^2}{1-c}& 0&0&2\hat{V}_{c}^2\\
	0& 0&2\frac{((\hat{s}_c+1)^2+c-1)}{1-c}&2(\hat{R}_b-\hat{R}_{GMV})&-2(\hat{R}_b-\hat{R}_{GMV})^2\\
	\hat{V}_{c}&0&2(\hat{R}_b-\hat{R}_{GMV})&\hat{V}_b&0\\
	-2\hat{V}_{c}(\hat{R}_b-\hat{R}_{GMV})&2\hat{V}_{c}^2&-2(\hat{R}_b-\hat{R}_{GMV})^2&0&2\hat{V}_b^2\\
	\end{pmatrix}.
	\end{equation}

\vspace{-0.32cm}
\begin{remark}
	In the case of the investor who invests into the GMV portfolio ($\gamma=\infty$), the formulas \eqref{alpha_opt} and \eqref{alpha_est} simplify to
 	\[
		\alpha^{*} = \frac{(1-c)(V_b-V_{GMV})}{c+(1-c)(V_b -V_{GMV})}
		\quad \text{and} \quad
		\hat{\alpha}_c^* = \frac{(1-c)(\hat{V}_b-\hat{V}_{c})}{c+(1-c)(\hat{V}_b-\hat{V}_{c})}.
		\]
	Moreover, the application of Theorem \ref{th1} leads to
	\begin{equation}
	\sqrt{n}(\hat{\alpha}_c^*- \alpha^*) \to \mathcal{N}\left( 0, \frac{2(1-c)c^2(L_{b}+1)}{((1-c)R_{\mathbf{b} }+c)^4}((2-c)L_{b}+c)\right)
	\end{equation}
	for $p/n\to c\in(0,1)$ as $n\to \infty$ with $L_b=V_b/V_{GMV}-1$, which coincides with the results obtained in Theorem 2 of \citet{BodnarDmytriv2019}.
\end{remark}

\subsection{Test based on a shrinkage estimator}\label{sec3b}

We use the properties of the shrinkage intensity $\alpha^*$ and of its consistent estimator $\hat{\alpha}_c^*$ to derive an asymptotic test on the structure of the EU portfolio. The testing hypotheses are given by
\begin{equation}\label{hypotheses_shrinkage}
H_{0}:\mathbf{w}_{EU}=\mathbf{w}_0\qquad \textrm{against} \qquad H_{1}:\mathbf{w}_{EU}\not= \mathbf{w}_0,
\end{equation}
which, in contrast to the hypotheses considered in Section~\ref{sec2}, allow to test the structure of the whole vector of the EU portfolio weights by using a single test avoiding the problem of multiplicity.

Following \citet{BodnarDmytriv2019}, the idea behind a statistical test based on the shrinkage approach is the usage $\mathbf{w}_0$ as a fixed target portfolio, i.e., to set $\mathbf{b} =\mathbf{w}_0$ in \eqref{hwGSE}. Since $\mathbf{w}_0$ is the EU optimal portfolio under the null hypothesis in \eqref{hypotheses_shrinkage}, its expected return and variance should satisfy
\begin{equation}\label{RV_w0}
R_{\mathbf{w}_0}=R_{GMV}+\gamma^{-1}s
\quad\text{and} \quad
V_{\mathbf{w}_0}=V_{GMV}+\gamma^{-2}s.
\end{equation}
As a result, the numerator in \eqref{alpha_opt} becomes
 \begin{equation*}\begin{split}
	A(\mathbf{w}_0)&=(R_{GMV}-R_b)\left( 1+ \frac{1}{1-c}\right)+\gamma(V_b-V_{GMV})+\frac{\gamma^{-1}}{1-c}s \\
	&=-\gamma^{-1}s\left( 1+ \frac{1}{1-c}\right)+\gamma^{-1}s+\frac{\gamma^{-1}}{1-c}s=0,
	\end{split}
	\end{equation*}
proving that
\begin{equation}\label{alpha_opt_H0}
\alpha^*=0 \quad \mbox{under} \quad H_0.
\end{equation}

Hence, for testing \eqref{hypotheses_shrinkage}, one can derive a test on the hypotheses
\begin{equation}\label{hypotheses_alpha}
H_{0}:\alpha^*(\mathbf{w}_0)=0\qquad \textrm{against} \qquad H_{1}:\alpha^*(\mathbf{w}_0) \not= 0,
\end{equation}
where the notation $\alpha^*(\mathbf{w}_0)$ denotes the optimal shrinkage intensity as in \eqref{alpha_opt} computed with target portfolio $\mathbf{w}_0$. It has to be noted that the hypotheses \eqref{hypotheses_shrinkage} and \eqref{hypotheses_alpha} are not equivalent. Nevertheless, the rejection of the null hypothesis in \eqref{hypotheses_alpha} ensures the rejection of the null hypothesis in \eqref{hypotheses_shrinkage} meaning that $\mathbf{w}_0$ is not the EU optimal portfolio.

Let $\hat{\alpha}_c^*(\mathbf{w}_0)$ be the consistent estimator of $\alpha^*(\mathbf{w}_0)$ as constructed in \eqref{alpha_est} when the shrinkage target is $\mathbf{b}=\mathbf{w}_0$. Then the application of Theorem \ref{th2} shows that
\[\hat{\alpha}_c^*(\mathbf{w}_0) \stackrel{a.s.}{\to} 0
\quad \mbox{for $\dfrac{p}{n} \to c\in[0,1)$ as $n \to \infty$},\]
when the null hypothesis in \eqref{hypotheses_shrinkage} is true.

Moreover, since the numerator in the expression of $\alpha^*(\mathbf{w}_0)$ in \eqref{alpha_opt} under the null hypothesis in \eqref{hypotheses_alpha} is equal to zero, i.e. $A=0$ where $A$ is defined before \eqref{dif_alp}, we get the following stochastic representation of $\sqrt{n}\hat{\alpha}_c^*(\mathbf{w}_0)$ expressed as
\begin{equation}
\sqrt{n}\hat{\alpha}_c^*(\mathbf{w}_0)=\frac{1}{\hat B_n}\mathbf{d}_0' \sqrt{n}\mathbf{t}
~ \text{ with }~
\mathbf{d}_0= \begin{pmatrix}
1+\frac{1}{1-c_n}\\
-\gamma \\
\frac{\gamma^{-1}}{1-c_n}\\
-1-\frac{1}{1-c_n}\\
\gamma \\
\end{pmatrix}
\end{equation}
and $\mathbf{t}$ is defined in \eqref{bt}. The application of Theorem \ref{th1} then leads to the following result

\begin{theorem}\label{th3}
	Assume that the conditions of Theorem \ref{th1} are fulfilled. Then, under the null hypothesis in \eqref{hypotheses_alpha}, it holds that
	\begin{equation}\label{asym_alpha_H0}
	\sqrt{n}\hat{\alpha}_c^*(\mathbf{w}_0)\stackrel{d}{\to} \mathcal{N}(0,C_{\alpha;0}),
	\end{equation}
	for $p/n \to c\in [0,1)$ as $n \to \infty$ with $C_{\alpha;0}=\frac{1}{B^2}\mathbf{d}_0'\mathbf{\Omega}_{\alpha} \mathbf{d}_0$ where $\mathbf{\Omega}_{\alpha} $ is given in \eqref{asym_bt_var} and $B$ is defined before \eqref{dif_alp}.
\end{theorem}

Replacing $B$ and $\mathbf{\Omega}_{\alpha}$ by their consistent estimators $\hat B_n^2$ and $\hat{\mathbf{\Omega}}_{\alpha;c}$, we get a consistent estimator for $C_{\alpha;0}$ expressed as
\begin{equation}\label{asym_sig_alp_c_H0}
\hat C_{\alpha;0}=\frac{1}{\hat B_n^2}\mathbf{d}_0'\hat{\mathbf{\Omega}}_{\alpha;c} \mathbf{d}_0\,.
\end{equation}
Then for testing hypotheses \eqref{hypotheses_alpha}, we obtain the following test statistic
\begin{equation}\label{T_alpha}
T_{\alpha}=\sqrt{n}\frac{\hat{\alpha}_c^*(\mathbf{w}_0)}{\sqrt{\hat C_{\alpha;0}}}=\sqrt{n}\frac{\hat{\alpha}_c^*(\mathbf{w}_0)\hat B_n}{\sqrt{\mathbf{d}_0'\hat{\mathbf{\Omega}}_{\alpha;c} \mathbf{d}_0}},
\end{equation}
where $\hat{\alpha}_c^*(\mathbf{w}_0)$ with $\mathbf{b}=\mathbf{w}_0$ and $\hat{\mathbf{\Omega}}_{\alpha;c}$ are given in \eqref{alpha_est} and \eqref{asym_sig_alp_c2}, respectively. Under the null hypothesis in \eqref{hypotheses_alpha} we get that
\[T_{\alpha}\stackrel{d}{\to} \mathcal{N}(0,1)\]
for $p/n \to c\in [0,1)$ as $n \to \infty$ and, hence, the hypothesis that $\mathbf{w}_0$ are the weights of the EU portfolio is rejected as soon as $|T_{\alpha}|>z_{1-\beta/2}$ where $z_{1-\beta/2}$ is the $(1-\beta/2)$ quantile of the standard normal distribution. Under the alternative hypothesis in \eqref{hypotheses_alpha}, the distribution of $\sqrt{n}\hat{\alpha}_c^*(\mathbf{w}_0)$ can still be well approximated by the normal distribution under the high-dimensional asymptotic regime and $\mathbf{d}_0'\hat{\mathbf{\Omega}}_{\alpha;c} \mathbf{d}_0$ provides a consistent estimator of its asymptotic variance. On the other side, it does not hold that $\hat{\alpha}_c^*(\mathbf{w}_0) \stackrel{a.s.}{\to} 0$ and consequently, the test based on $T_{\alpha}$ can detect the deviation in the null hypotheses of both \eqref{hypotheses_shrinkage} and \eqref{hypotheses_alpha}.

\begin{remark}
	Using that $s=\gamma(R_{\mathbf{w}_0}-R_{GMV})$ (see \eqref{RV_w0}) and $\hat R_{\mathbf{w}_0}$ and $\hat R_{GMV}$ are consistent estimators of $R_{\mathbf{w}_0}$ and $R_{GMV}$, respectively (see \eqref{consist_est}), another consistent estimator of $\mathbf{\Omega}_{\alpha}$ under $H_0$ in \eqref{hypotheses_alpha} is given by
	
		{\small \begin{equation}\label{asym_sig_alp_c3}
		\widetilde{\mathbf{\Omega}}_{\alpha;c}=
		\begin{pmatrix}
		\frac{\hat{V}_{c}(\gamma(\hat R_{\mathbf{w}_0}- \hat R_{GMV})+1)}{1-c} & 0 &0&\hat{V}_{c}&-2\hat{V}_{c}(\hat{R}_{\mathbf{w}_0}-\hat{R}_{GMV})\\
		0 & 2\frac{\hat{V}_{c}^2}{1-c}& 0&0&2\hat{V}_{c}^2\\
		0& 0&2\frac{((\gamma(\hat R_{\mathbf{w}_0}- \hat R_{GMV})+1)^2+c-1)}{1-c}&2(\hat{R}_{\mathbf{w}_0}-\hat{R}_{GMV})&-2(\hat{R}_{\mathbf{w}_0}-\hat{R}_{GMV})^2\\
		\hat{V}_{c}&0&2(\hat{R}_{\mathbf{w}_0}-\hat{R}_{GMV})&\hat{V}_{\mathbf{w}_0}&0\\
		-2\hat{V}_{c}(\hat{R}_{\mathbf{w}_0}-\hat{R}_{GMV})&2\hat{V}_{c}^2&-2(\hat{R}_{\mathbf{w}_0}-\hat{R}_{GMV})^2&0&2\hat{V}_{\mathbf{w}_0}^2\\
		\end{pmatrix}.
		\end{equation}}

	Then, the hypotheses in \eqref{hypotheses_alpha} can also be tested by using the following test statistic
	\begin{equation}\label{tT_alpha}
	\widetilde T_{\alpha}=\sqrt{n}\frac{\hat{\alpha}_c^*(\mathbf{w}_0) \hat{B}_n}{\sqrt{\mathbf{d}_0'\widetilde{\mathbf{\Omega}}_{\alpha;c} \mathbf{d}_0}}
	\end{equation}
	which is asymptotically standard normally distributed under $H_0$ in \eqref{hypotheses_alpha}.
\end{remark}

\begin{remark}
	Using the duality between the test theory and confidence interval (see, \citet{aitchison1964confidence}), the null hypothesis in \eqref{hypotheses_alpha} and consequently in \eqref{hypotheses_shrinkage} are rejected at significance level $\beta$ as soon as the $(1-\beta)$ confidence interval constructed for $\alpha^*(\mathbf{w}_0)$ does not include zero. This confidence interval in the case of the test $T_{\alpha}$ has the boundaries
	\begin{equation}\label{CI-T_alp}
	\hat{\alpha}_c^*(\mathbf{w}_0)\pm\frac{z_{1-\beta/2}}{\sqrt{n}}\frac{\sqrt{\mathbf{d}_0'\hat{\mathbf{\Omega}}_{\alpha;c} \mathbf{d}_0}}{\hat{B}_n},
	\end{equation}
	while for the test based on $\widetilde{T}_{\alpha}$ we get
	\begin{equation}\label{CI-tT_alp}
	\hat{\alpha}_c^*(\mathbf{w}_0)\pm\frac{z_{1-\beta/2}}{\sqrt{n}}\frac{\sqrt{\mathbf{d}_0'\widetilde{\mathbf{\Omega}}_{\alpha;c} \mathbf{d}_0}}{\hat{B}_n}.
	\end{equation}
\end{remark}

\begin{figure}[!h]
	\centering
	\begin{tabular}{ll}
			\includegraphics[scale=0.45]{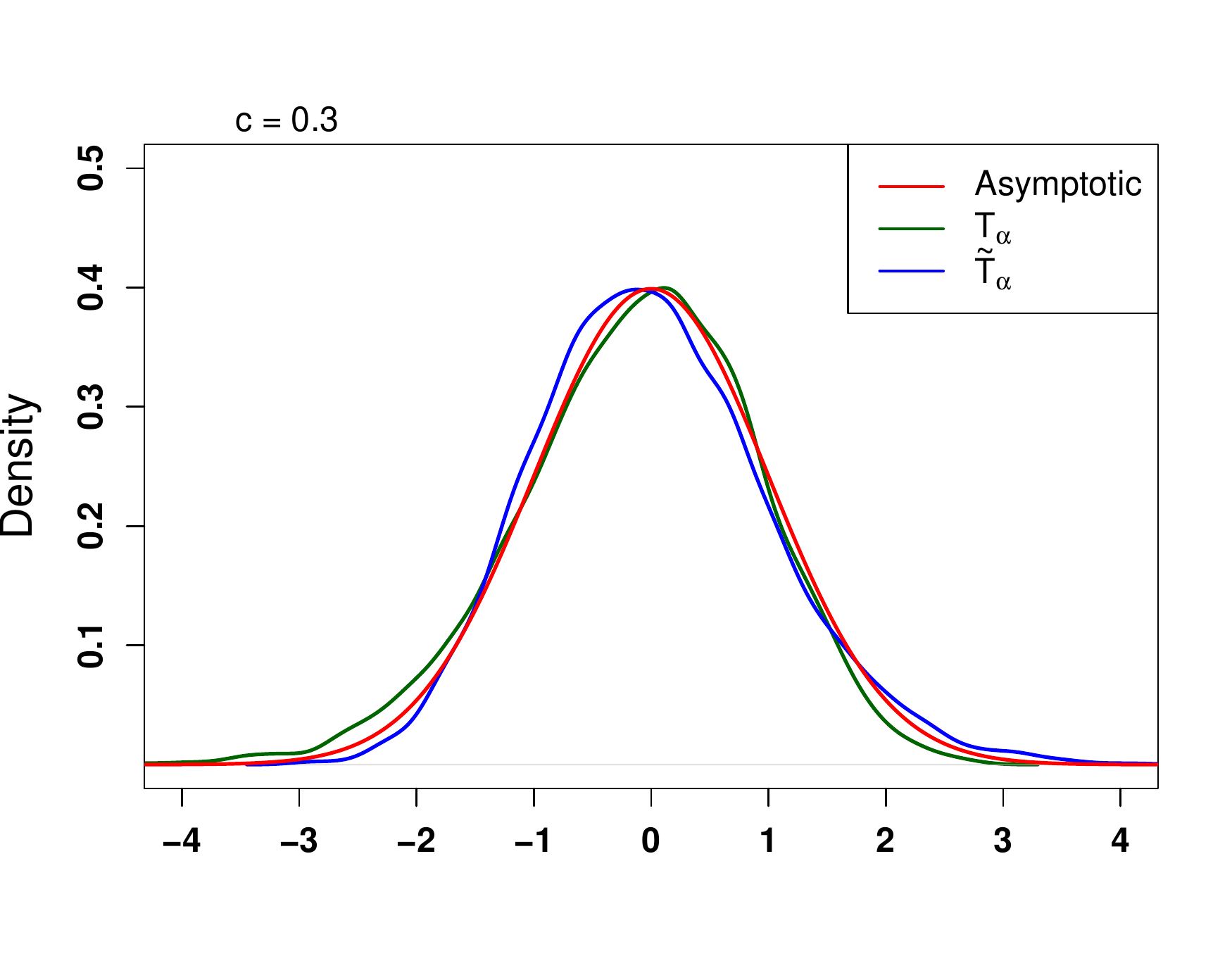}& 
		\includegraphics[scale=0.45]{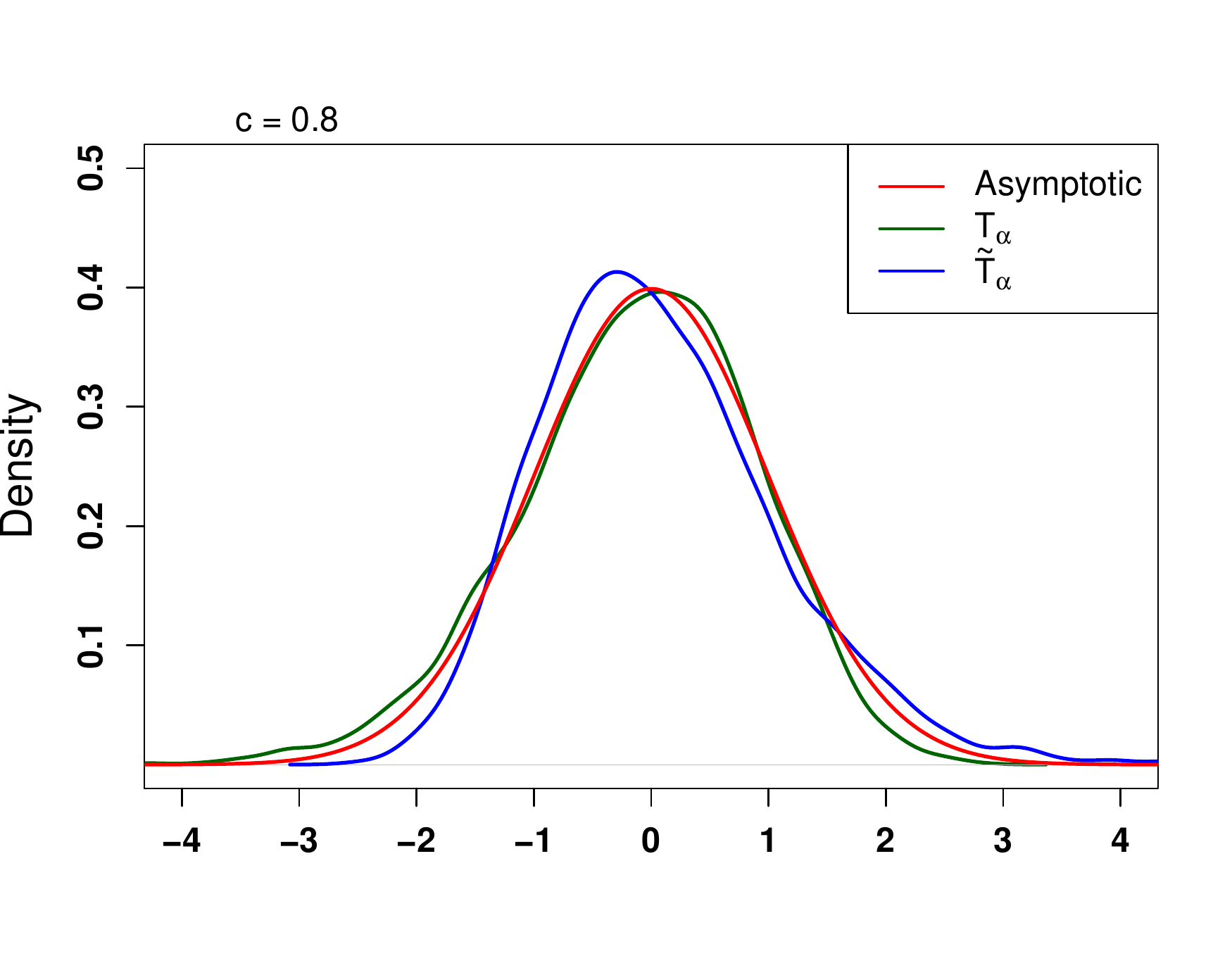}
	\end{tabular}
\vspace{-0.5cm}
	\caption{{The high-dimensional asymptotic normal approximation of the densities of $T_{\alpha}$ and $\widetilde{T}_{\alpha}$ together with their kernel density estimators for $\gamma=5$, $p=300$ and $c_n=p/n \in\{0.3,0.8\}$.}} \label{table: tab2}
	\label{fig2}
\end{figure}

To assess the precision of the asymptotic distribution we use a similar setting as in the last section. In Figure \ref{fig2} we show the KDEs of the distribution of the test statistics $T_{\alpha}$ and $\widetilde{T}_{\alpha}$ under the null hypothesis together with their  high-dimensional asymptotic distribution. The latter approximates the simulated exact distributions very precisely, although the the fit appears to be slightly better for $T_{\alpha}$. The empirical size on both cases is close to the nominal size of $5\%$ as it is shown in Table \ref{table: tab2}. Summarizing,  we conclude that the high-dimensional asymptotic distribution provide a good approximation for proposed test statistics for different values of $c$.
\begin{table}[!h]
	\centering		
	\begin{tabular}{lcc}
			\hline \vspace{0.05cm}
			& $\mathbf{c=0.3}$ & $\mathbf{c=0.8}$ \\ \hline \hline \vspace{0.05cm}
			$T_{\alpha}$	& 0.048 & 0.054 \\ \cline{2-3} \vspace{0.05cm}
			$\tilde{T}_{\alpha}$	& 0.052 & 0.053 \\ \hline
	\end{tabular}	
	\caption{Empirical sizes of the two tests based on $T_{\alpha}$ and $\tilde{T}_{\alpha}$ using $5 \cdot 10^3$  replications.} \label{table: tab2}
\end{table}

\section{Simulation and empirical study}\label{sec4}
The performance of the derived test is investigated throughout an extensive simulation study. In particular, we explore the behavior of the test with respect to its power characteristics and receiver operative characteristic curves. Additionally, we apply the derived inference procedure to the real data in this section.
\subsection{Simulation study}\label{sec4a}

The sample of asset returns $\mathbf{x}_1,\mathbf{x}_2, \ldots,\mathbf{x_n}$ are generated independently from $\mathcal{N}_p(\boldsymbol{\mu}, \mathbf{\Sigma})$. To mimic the bahavior of real data we generate the eigenvalues of population covariance matrix $\mathbf{\Sigma}$ according to the law $\lambda_i=0.1 e^{\delta c(i-1)/p}$, $i=1,\ldots p$ (see, \citet{bodnar2016okhrin}) and take its eigenvectors from the spectral decomposition of the standard Wishart random matrix. Then, the covariance matrix is given as follows
\begin{equation}
\mathbf{\Sigma}=\Theta \mathbf{\Lambda} \Theta',
\end{equation}
where $\mathbf{\Lambda}$ is a diagonal matrix of the predefined eigenvalues and $\Theta$ is a $p\times p$ matrix of eigenvectors. By changing the value of $\delta$, we can control the conditional index of the covariance matrix for different values of $c$. We set condition index equals to 450. This setting reflects the parametrisation we observed in the empirical study in the next section. The mean vector is randomly generated from $\mathit{U}(-0.2, 0.2)$, which also corresponds to the natural behavior of daily asset returns.

We assume that the portfolio weights and thus the shrinkage intensity change due to a change in the mean of asset returns. Under the alternative hypothesis, there is an additive shift to the mean vector of the asset returns defined as
\begin{equation}\label{scenario1}
\boldsymbol{\mu}_1=\boldsymbol{\mu} + \mathbf{\epsilon},
\end{equation}
where
\begin{equation*}
\mathbf{\epsilon}=-a\cdot( \underbrace{1, \ldots, 1}_{m}, \underbrace{0, \ldots, 0}_{m}) ,
\end{equation*}
where $a=0.01\kappa$,\, $\kappa \in \{0, 1, 2,\ldots, 35\}$, $m =0.5p$. Thus we assume that the expected return on the assets with high variance decreases.

\begin{figure}[th!]
	\vspace{-0.3cm}
	\centering
	\begin{tabular}{ll}
		\includegraphics[width=0.48\linewidth]{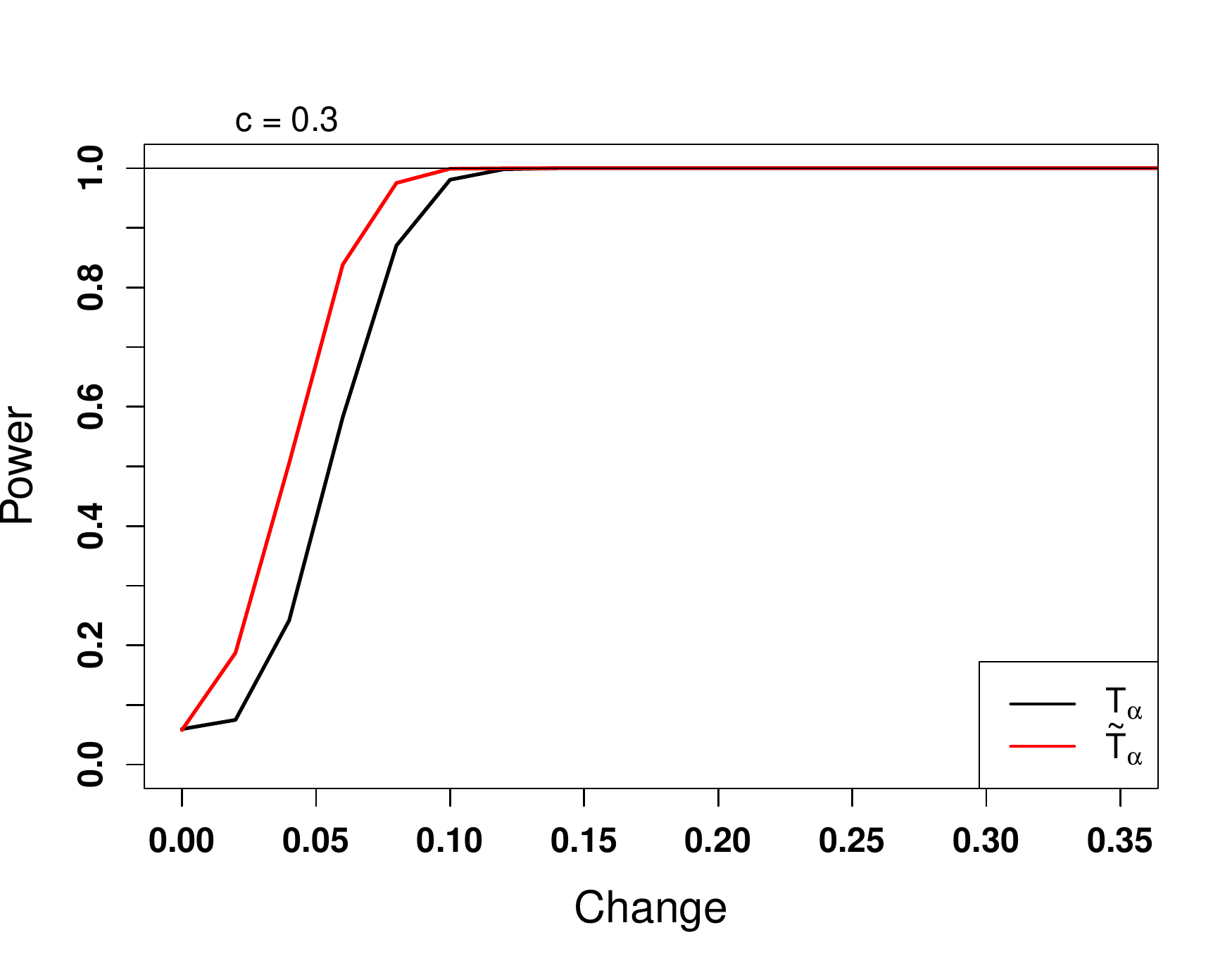} &\hspace{-0.3cm}	\includegraphics[width=0.48\linewidth]{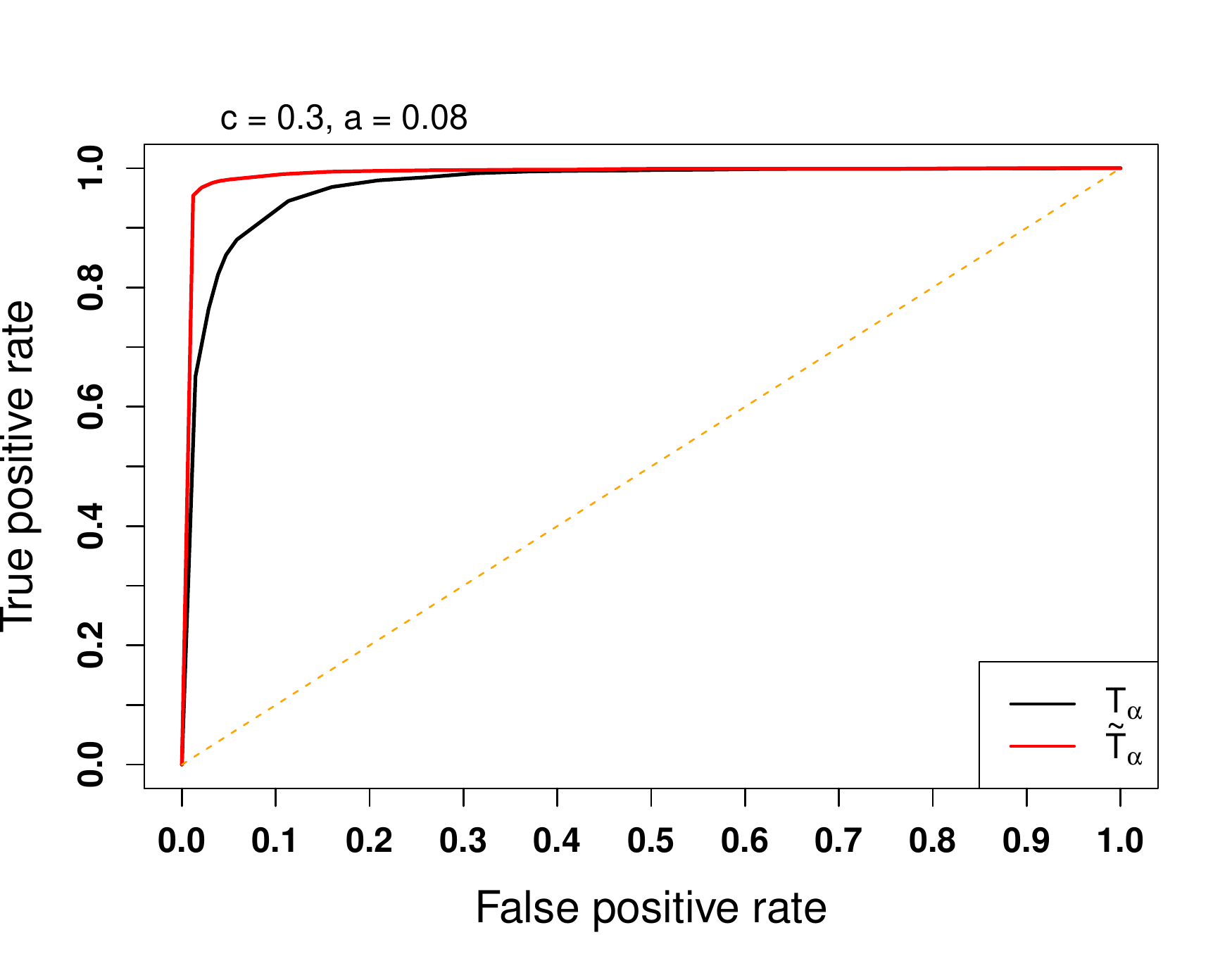}\\
		\includegraphics[width=0.48\linewidth]{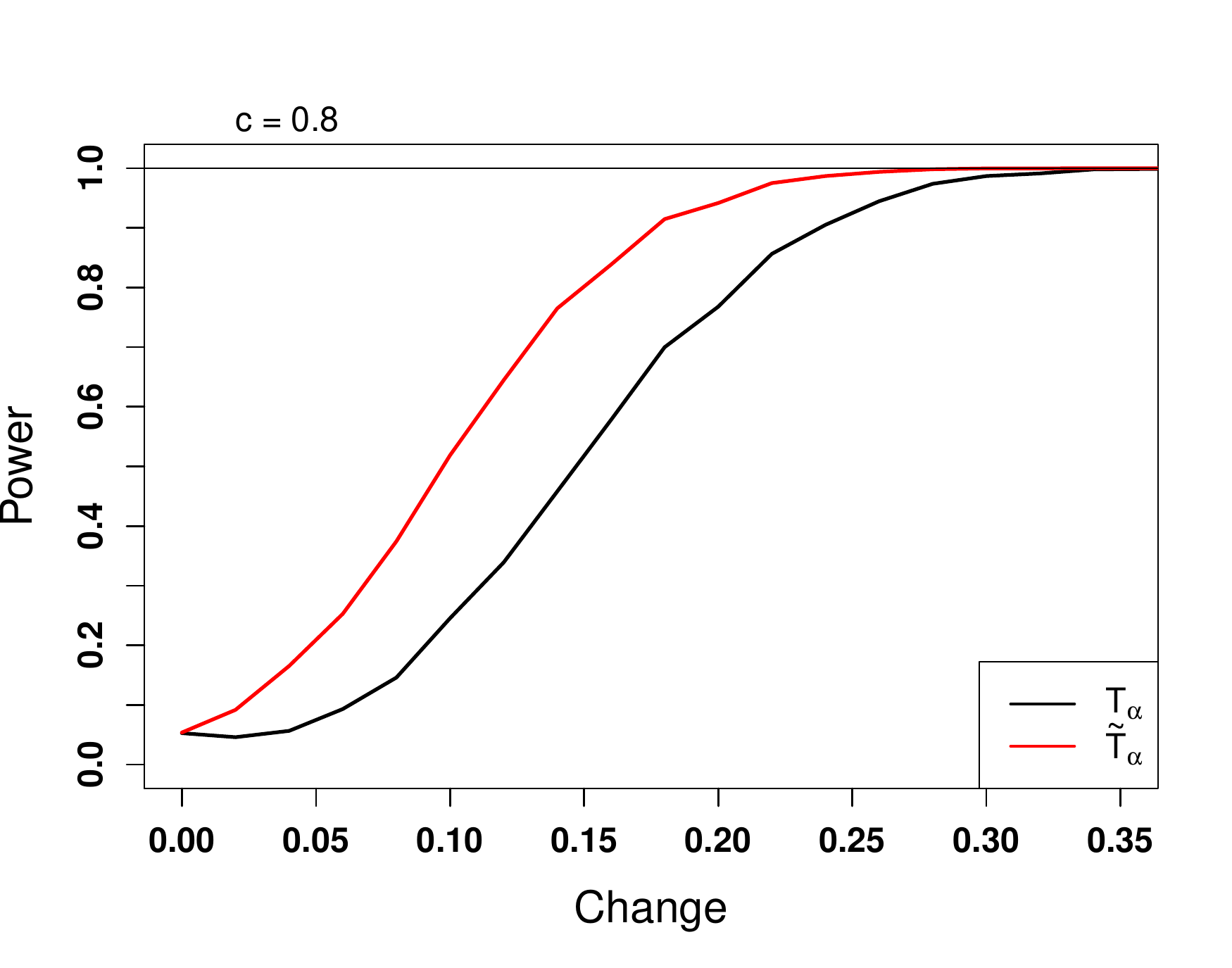} &\hspace{-0.3cm}	\includegraphics[width=0.48\linewidth]{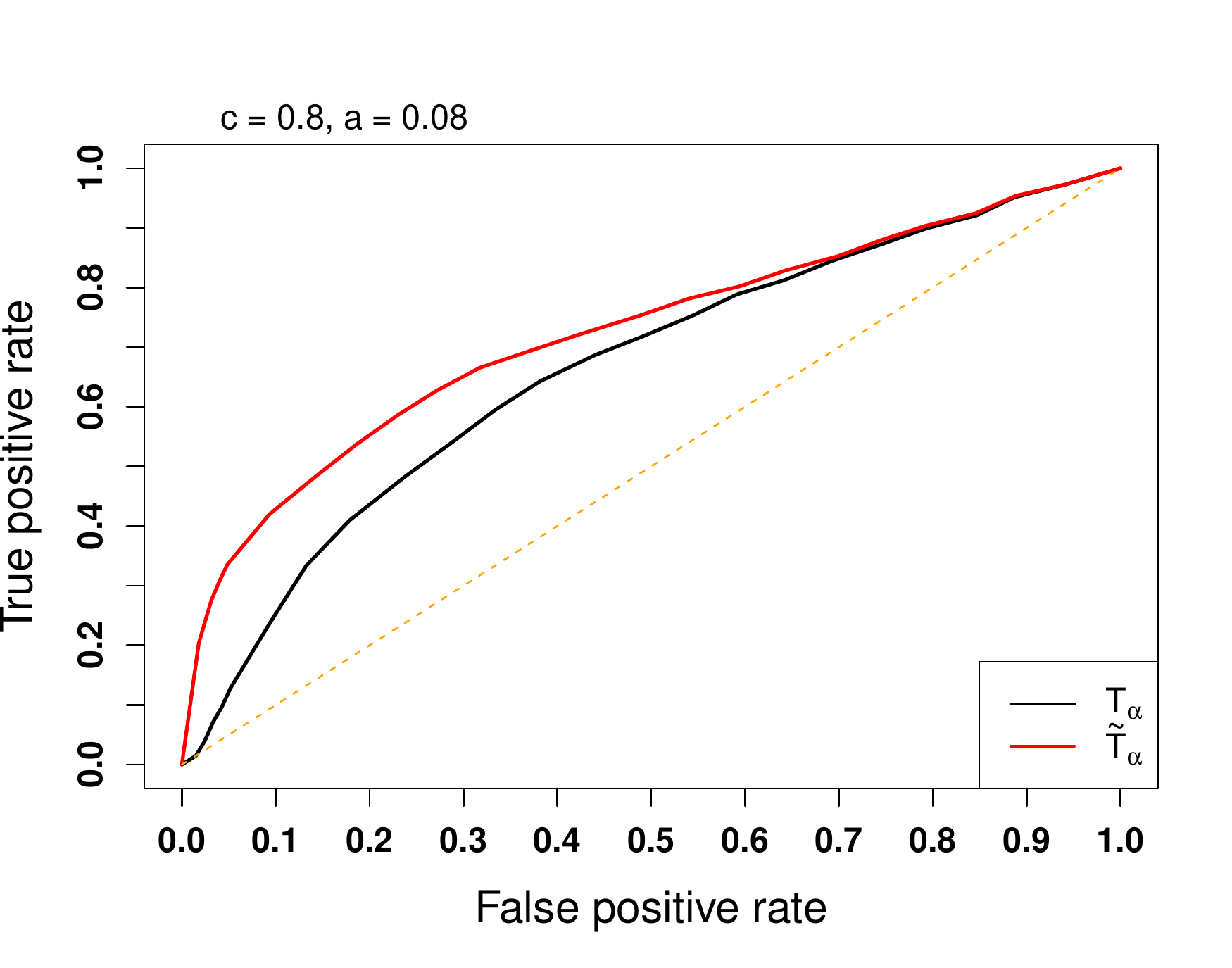}\\	
	\end{tabular}
	\caption{Empirical power functions of the proposed tests as a function of the change  $a$ (left) and ROC curves of two tests for $a=0.08$ (right) for different values of $c$ according to the scenario given in (\ref{scenario1}) and $p=300$.}
	\label{fig:fig3}	
\end{figure}

We conduct the test at the significance level $\alpha = 0.05$. We put $p=300$ and $c\in \{0.3, 0.8\}$. The number of repetitions is $10^5$ and $\gamma=5$. For the ROC curves we fix $a$ at $0.08$. The results are illustrated in  Figure \ref{fig:fig3}. It can be seen that both tests display an overall consistency and a good performance in terms of power functions and ROC curves. The behavior is better for smaller values of $c$ and not substantially worse in case of $c=0.8$. The test based on the test statistic given in \eqref{tT_alpha} outperforms the test given in \eqref{T_alpha} and demonstrates a satisfactory power.

%\begin{figure}[th!]
%	\centering
%	\begin{tabular}{cc}
%		\includegraphics[width=0.46\linewidth]{Hist_03_p=100.pdf}&	\includegraphics[width=0.46\linewidth]{Hist_03_p=300.pdf}\\	
%		\includegraphics[width=0.46\linewidth]{Hist_06_p=100.pdf}&
%		\includegraphics[width=0.46\linewidth]{Hist_06_p=300.pdf}\\	\includegraphics[width=0.46\linewidth]{Hist_08_p=100.pdf}&	
%	\includegraphics[width=0.46\linewidth]{Hist_08_p=300.pdf}
%	\end{tabular}
%	\caption{Kernel vs asymptotic density of the estimated shrinkage intensity for different values of $c$, $c\in\{0.3, 0.5, 0.6, 0.8\}$ with $p=100$(left) and $p=300$(right).}
%	\label{fig:fig3}	
%\end{figure}
%	

\begin{figure*}[th!]
	\centering
	\begin{tabular}{cc}
		\includegraphics[width=0.48\linewidth]{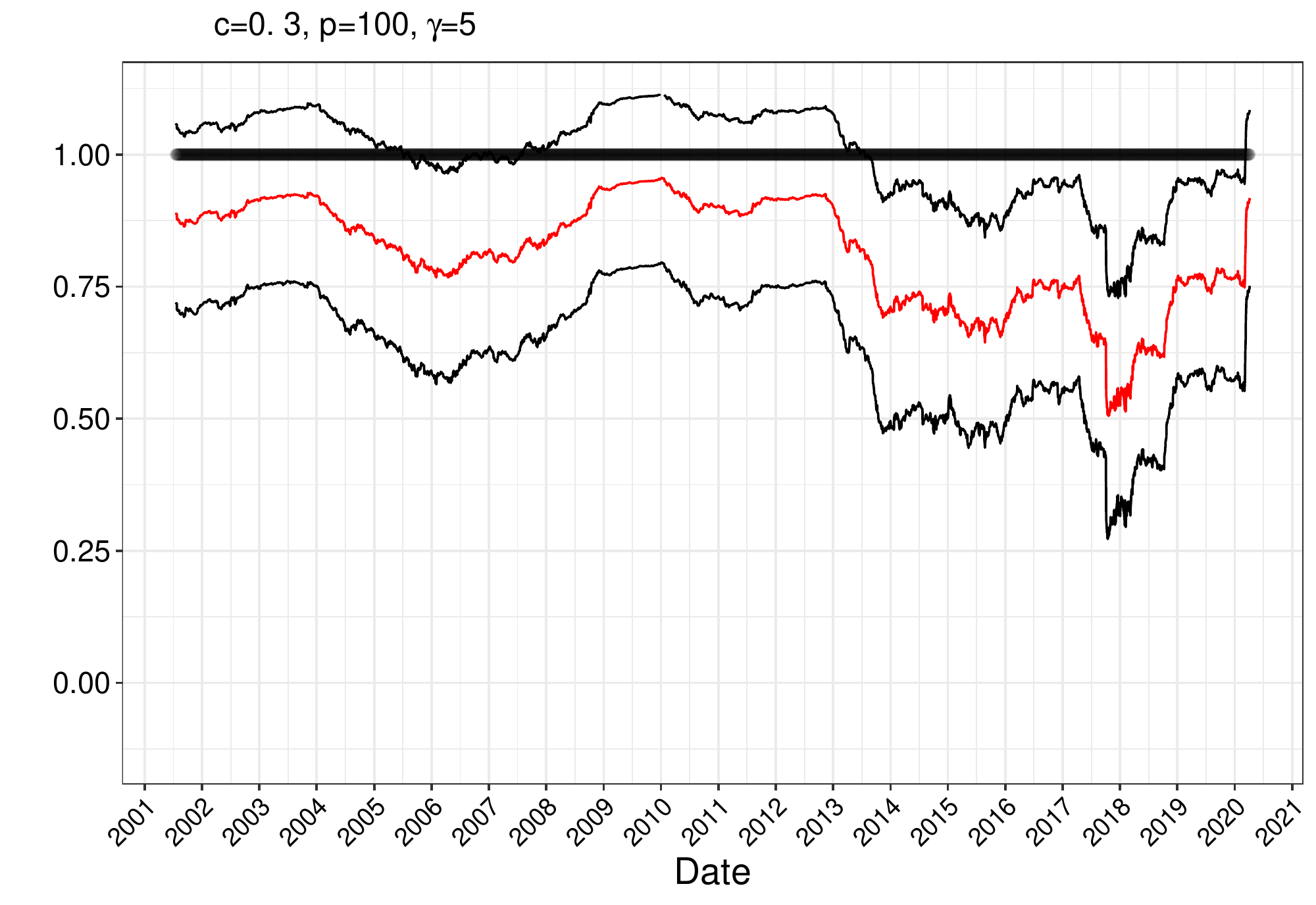} &	\includegraphics[width=0.48\linewidth]{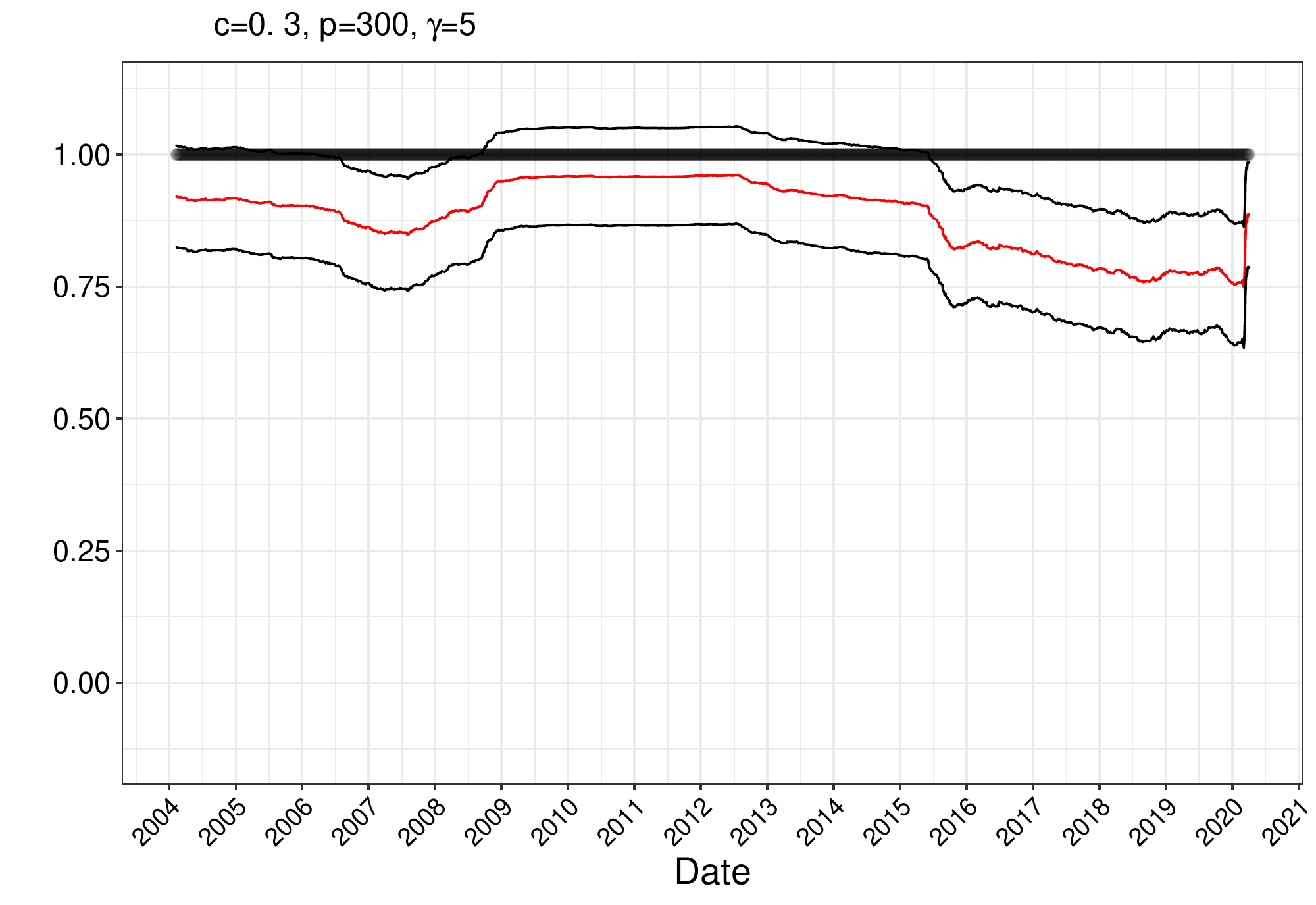}\\	
		
		\includegraphics[width=0.48\linewidth]{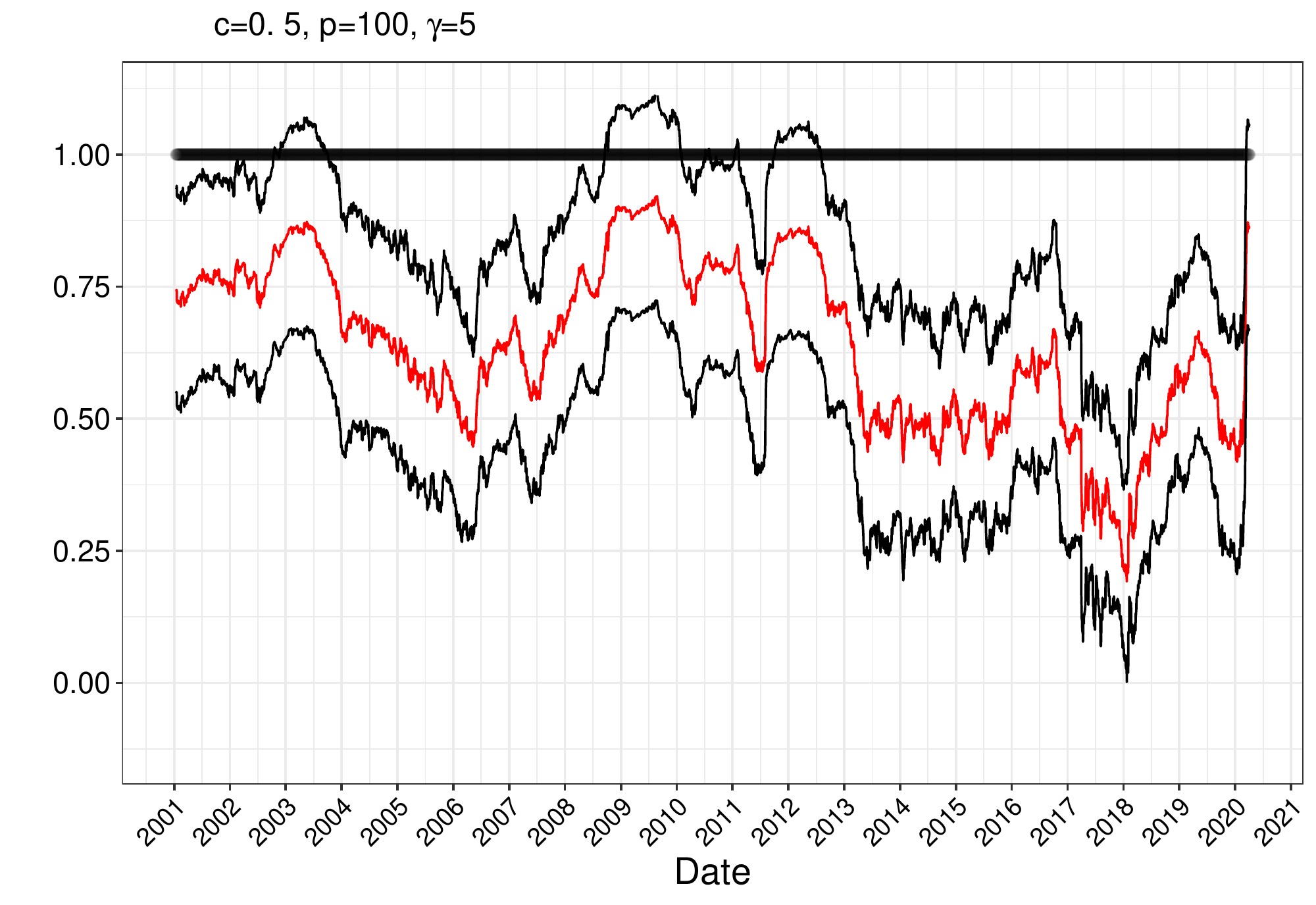} &	\includegraphics[width=0.48\linewidth]{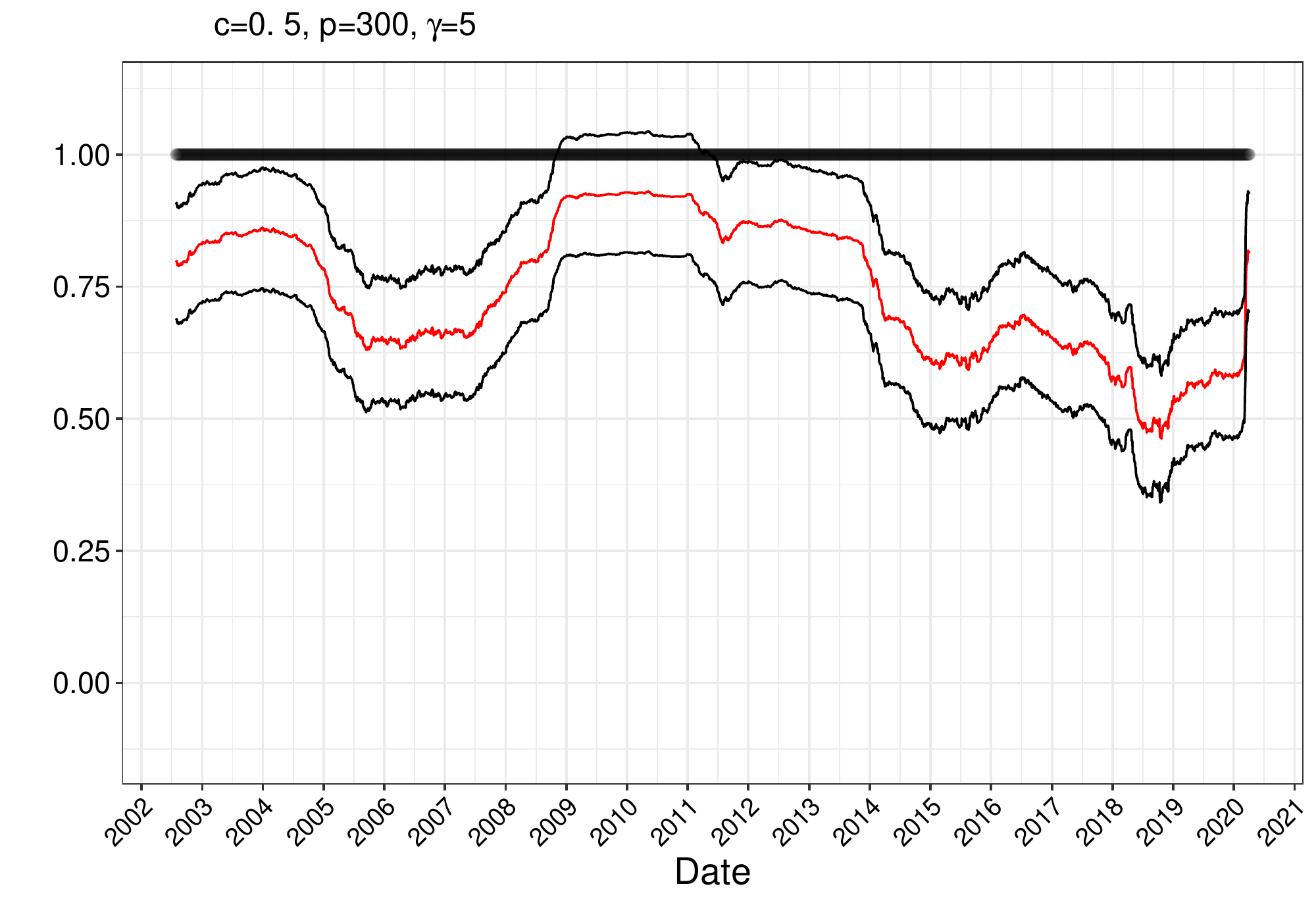}\\
		
		\includegraphics[width=0.48\linewidth]{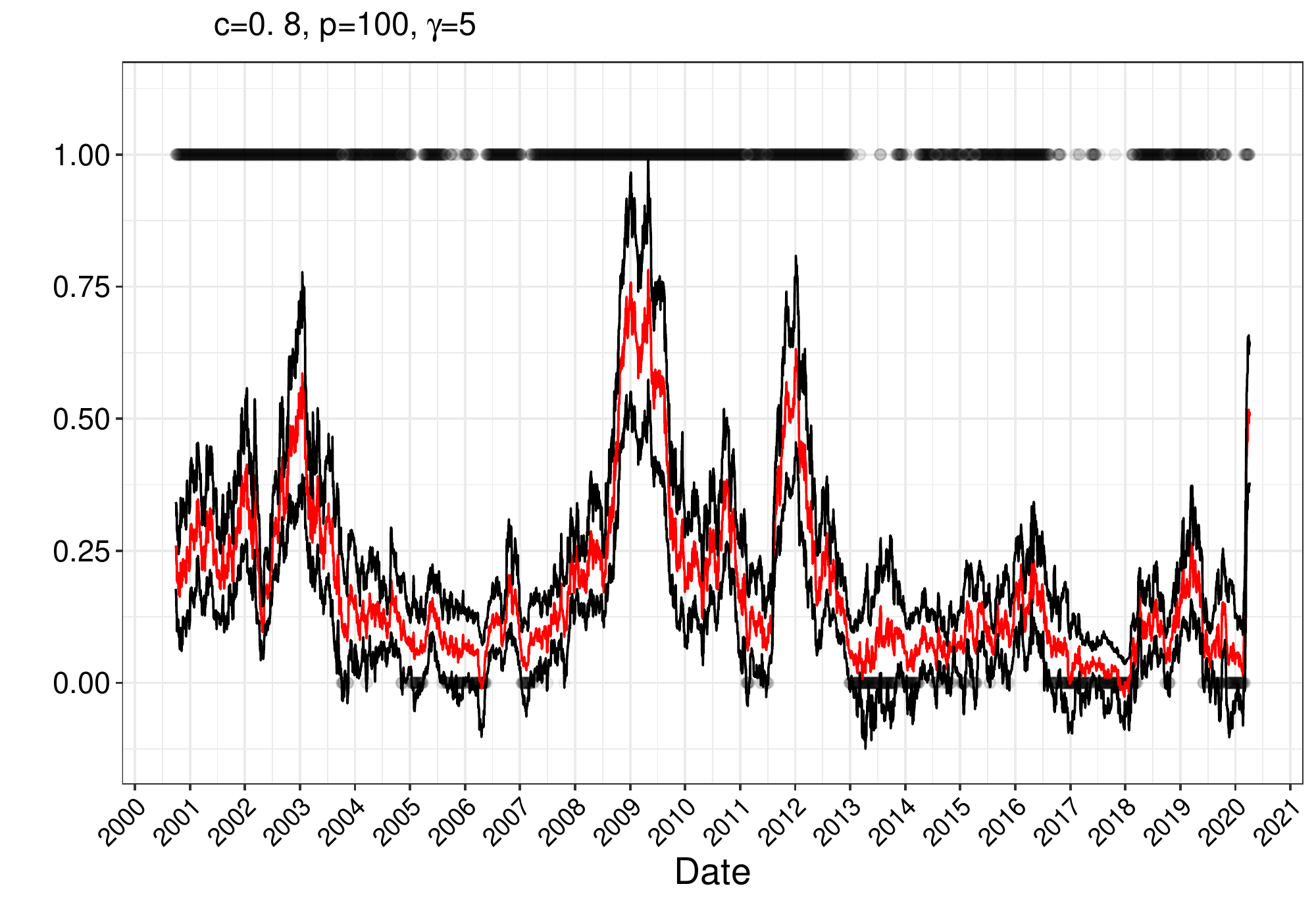} &	\includegraphics[width=0.48\linewidth]{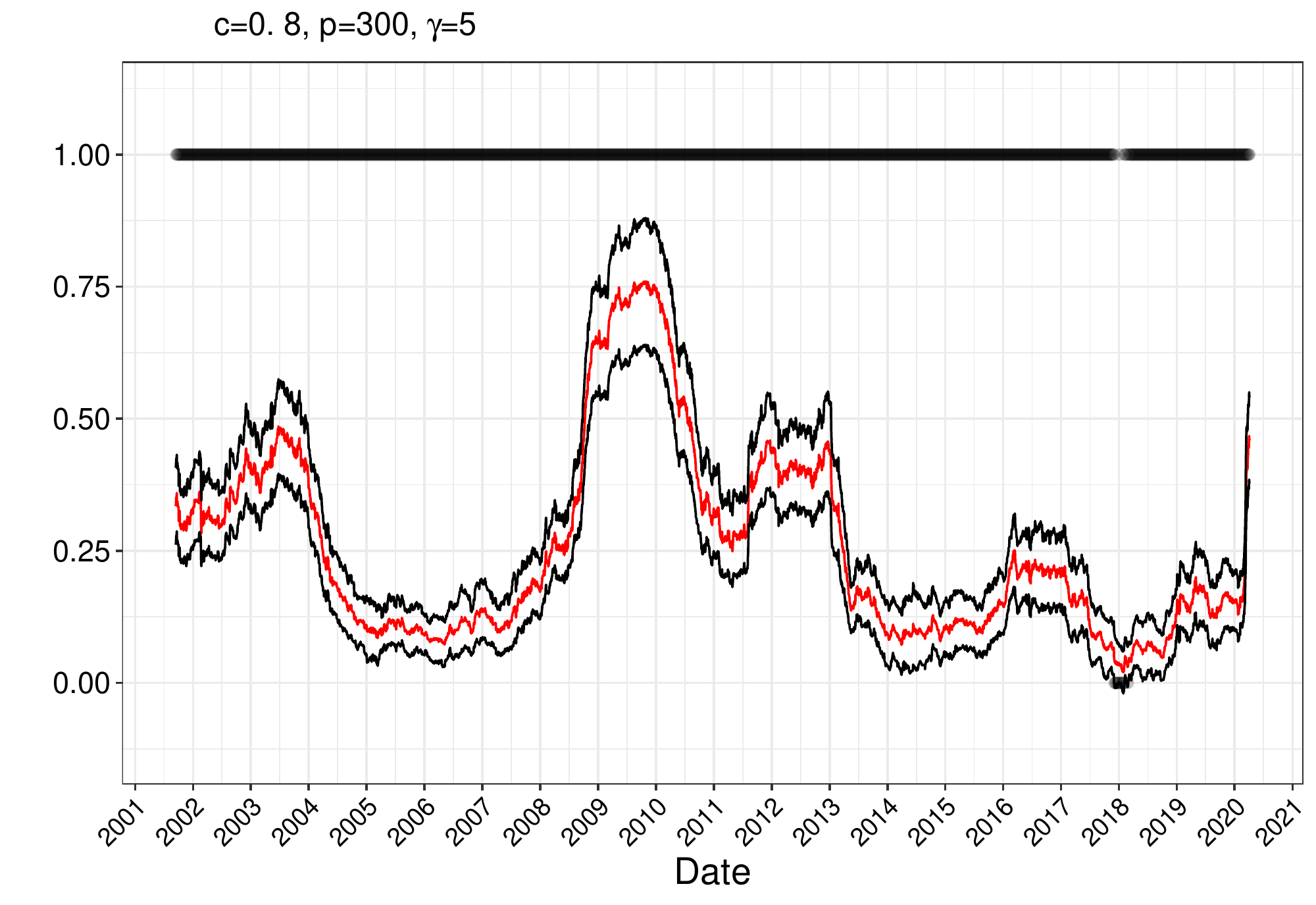}\\
	\end{tabular}
	\caption{Estimated shrinkage intensities for the equally weighted portfolio as the target portfolio ($p=100$ on the right and $p=300$ on the left) with 95\% pointwise confidence intervals. The black dots indicate the periods with rejected  $H_0$ (1-values) and not rejected  $H_0$ (0- values).}
	\label{fig:fig5}	
\end{figure*}

\subsection{Empirical study}\label{sec4b}
In this section, we apply the derived theoretical results to  real data. The objective is to determine the periods where the shrinkage intensity is significantly different from zero and thus the EU optimal portfolio is significantly different from the target or the benchmark portfolio $\bm{b}$.  This study is based on daily return data of all companies listed in the S\&P 500 index for the period from April 1999 to March 2020. We assume that the investor allocates her wealth to portfolios of size $p \in \{100, 300\}$ with daily reallocation. She selects the first $p$ assets in alphabetic order from the available data. The sample size $n$ is chosen to attain $c\in \{0.3, 0.5, 0.8\}$, i.e. $n=p/c$. We put $\gamma=5$ which is a common value for the risk aversion coefficient in financial literature. As the target portfolio we consider the equally weighted portfolio with all weights equal $1/p$. Despite of its simplicity this portfolio appears to show a superior long-run performance and dominates many more sophisticated trading strategies (see \cite{deMiguel1n2019}).

Figure \ref{fig:fig5} shows the time series of estimated shrinkage intensities together with 95\% confidence intervals as defined in (\ref{CI-tT_alp}).
If $c=0.3$, then the shrinkage intensity is close to one indicating that the EU portfolio clearly dominates both benchmarks in the convex combination. This is due to the fact, that the investor has more historical data to estimate the unknown parameters and the estimation risk is relatively low. If $c$ increases, the sample available for a portfolio of a fixed size gets smaller and the shrinkage intensity shifts towards zero. The benchmark portfolio gets higher weight and for $c=0.8$ it even becomes dominant. The same reasoning applies if we analyse the impact of increase in $p$ from 100 to 300. Fixed $c$ and larger $p$ increase the sample size $n$ and has a stabilizing impact on the shrinkage intensity.

We cannot reject the null hypothesis of the test based on $\tilde{T}_\alpha$ in (\ref{tT_alpha}) that the shrinkage intensity is zero if the confidence intervals cover the zero value (see Remark 3 above). The figures reveal that we never opt for $H_0$ if $c=0.3$ or $0.5$. Thus for this parameter constellation the portfolio weights of the EU portfolio are always significantly different from the weights of the equally weighted portfolio. The situation changes for $c=0.8$ where we do have periods with not rejected $H_0$ in (\ref{hypotheses_alpha}). Similar behavior is observed for $p=300$ too, however, here the intensities and their variances are more stable leading to less periods with not rejected $H_0$.
\begin{figure}[th!]
	\centering
	\includegraphics[width=0.8\linewidth]{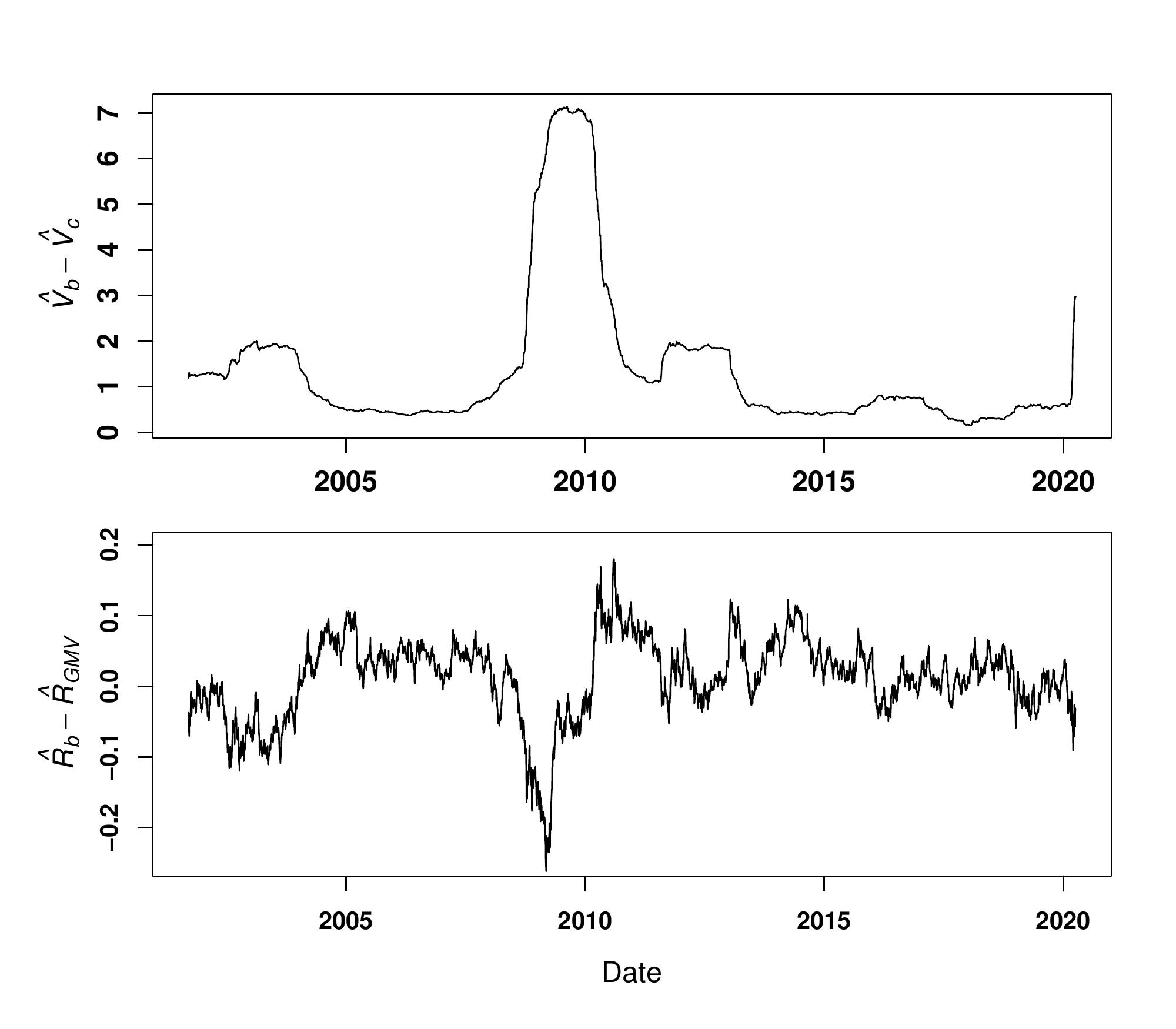}
	\caption{Components of the estimated shrinkage intensity given in \eqref{alpha_est} using equally weighted target for $c= 0.8$, $p=300$ and $\gamma=5$.}
	\label{fig:figRVs}	
\end{figure}

Recall that a non-rejection of $H_0$  in (\ref{hypotheses_alpha}) does not guarantee that the weights of the EU portfolio coincide with the weights of the target portfolio. To elaborate on the difference between the two portfolios  and to get more economic insight into the dynamics of the intensities  we consider Figure \ref{fig:figRVs}. Here we plot the difference between the means and variances of the GMV and the equally weighted benchmark. These quantities determine the behavior of the empirical shrinkage intensity in (\ref{alpha_opt}). On the one hand, we observe in Figure \ref{fig:fig5} that the shrinkage intensity increases during a crisis period, e.g. 2002-2003 and 2008-2010. This seems to be surprising since the volatility of returns is high in this period and  the equally weighted  portfolio is believed to reduce the risk. However, Figure \ref{fig:figRVs} shows that the variance of the benchmark portfolio is much higher (i.e. $\hat{V}_b>\hat{V}_c$)  and its return is much lower (i.e. $\hat{R}_b<\hat{R}_{GMV}$) compared to the GMV portfolio in the crisis period  leading to a higher relative precision and efficiency of the EU portfolio. On another hand, the mean returns and the variances are almost indistinguishable in calm periods leading to shrinkage intensities closer to zero and even insignificant for larger $c$'s. Thus we conclude that non-rejecting $H_0$ is driven by high similarity between the mean and the variance of the target and GMV portfolios.
% This explains the conclusions from the tests.
%We employ a daily rebalancing strategy, which is pretty common in practice.
%Within each rolling-window step, the covariance matrix of asset returns for the
%investment day $t$ is estimated at the end of day t?1 using approximately
%the most recent n daily in-sample observations.

\section{Summary}\label{sec5}
This paper is dedicated to portfolio selection problems driven by high-dimensional financial data sets. In particular, we deal with optimal asset allocation in a high-dimensional asymptotic regime, namely when the number of assets and the sample size tend to infinity at the same rate. Due to the curse of dimensionality in the parameter estimation process, asset allocation for such portfolios becomes a challenging task. Using the techniques from the theory of random matrices, new inferential procedures based on the optimal shrinkage intensity for testing the efficiency of the high-dimensional EU portfolio are developed and the asymptotic distributions of the proposed test statistics are derived. In extensive simulations, we show that the suggested tests have excellent performance characteristics for various values of $c$. The practical advantage of the proposed procedures are demonstrated in en empirical study based on stocks included into the S\&P 500 index.

{\footnotesize
\bibliography{bibliography}
%\nocite{*}
}

\section{Appendix}\label{sec-app}
In this section the proofs of the theoretical results are given. The proof of Theorem \ref{th1} is based on Lemmas \ref{lem1}-\ref{lem2}.

\begin{lemma}\label{lem1}
	Let $\mathbf{z}_1$, ..., $\mathbf{z}_n$ be an independent sample from the $p$-dimensional standard normal distribution and let
	\begin{equation}
	\mathbf{S}_n=\frac{1}{n-1} \sum_{j=1}^n (\mathbf{z}_j-\mathbf{\bar{z}})(\mathbf{z}_j-\mathbf{\bar{z}})'
	\end{equation}
	be the corresponding sample covariance matrix. Let $\mathbf{m}_1$, $\mathbf{m}_2$, and $\mathbf{m}_3$ be the $p$-dimensional vector of constants with the  Euclidean norms equal to one. Then
	\begin{equation} \begin{split}
		\label{lem1_eq1}
		\sqrt{n}\left( \begin{matrix}
		\mathbf{m}_1'\mathbf{S}_n\mathbf{m}_1-1\\
		\mathbf{m}_2'\mathbf{S}_n^{-1}\mathbf{m}_2-\frac{1}{1-c_n}\\
		\mathbf{m}_2'\mathbf{S}_n^{-1}\mathbf{m}_3-\frac{1}{1-c_n}\mathbf{m}_2'\mathbf{m}_3\\
		\mathbf{m}_3'\mathbf{S}_n^{-1}\mathbf{m}_3-\frac{1}{1-c_n}
		\end{matrix}\right)
		\stackrel{d}{\to}\mathcal{N}_4\left(\mathbf{0} , \frac{2}{c}\boldsymbol{\Theta}(\mathbf{m}_1,\mathbf{m}_2,\mathbf{m}_3)\circ\boldsymbol{\Lambda} \right),
		\end{split} 
	\end{equation}
	with
\begin{equation}\label{bTheta}
\boldsymbol{\Theta}(\mathbf{m}_1,\mathbf{m}_2,\mathbf{m}_3)=\left(\begin{matrix}
1 & \lim\limits_{n\to\infty}(\mathbf{m}_1'\mathbf{m}_2)^2&
\lim\limits_{n\to\infty}(\mathbf{m}_1'\mathbf{m}_2)(\mathbf{m}_1'\mathbf{m}_3)&\lim\limits_{n\to\infty}(\mathbf{m}_1'\mathbf{m}_3)^2\\
\lim\limits_{n\to\infty}(\mathbf{m}_1'\mathbf{m}_2)^2&1 &
\lim\limits_{n\to\infty}(\mathbf{m}_2'\mathbf{m}_3)&\lim\limits_{n\to\infty}(\mathbf{m}_2'\mathbf{m}_3)^2\\
\lim\limits_{n\to\infty}(\mathbf{m}_1'\mathbf{m}_2)(\mathbf{m}_1'\mathbf{m}_3)&
\lim\limits_{n\to\infty}(\mathbf{m}_2'\mathbf{m}_3)&0.5+0.5 \lim\limits_{n\to\infty}(\mathbf{m}_2'\mathbf{m}_3)^2 &\lim\limits_{n\to\infty}(\mathbf{m}_2'\mathbf{m}_3)\\
\lim\limits_{n\to\infty}(\mathbf{m}_1'\mathbf{m}_3)^2& \lim\limits_{n\to\infty}(\mathbf{m}_2'\mathbf{m}_3)^2 &\lim\limits_{n\to\infty}(\mathbf{m}_2'\mathbf{m}_3) & 1\\
\end{matrix} \right)
\vspace{0.4cm}
\end{equation}
	and
	\begin{equation}\label{bLambda}
	\boldsymbol{\Lambda}=  \left(\begin{matrix}
	c & -\frac{c}{1-c}& -\frac{c}{1-c}& -\frac{c}{1-c} \\
	-\frac{c}{1-c} & \frac{c}{(1-c)^3}&\frac{c}{(1-c)^3}&\frac{c}{(1-c)^3}\\
	-\frac{c}{1-c} &\frac{c}{(1-c)^3}&\frac{c}{(1-c)^3}&\frac{c}{(1-c)^3}\\
	-\frac{c}{1-c}&\frac{c}{(1-c)^3}&\frac{c}{(1-c)^3}&\frac{c}{(1-c)^3}\\
	\end{matrix} \right),
		\vspace{0.4cm}
	\end{equation}

	where the symbol $\circ$ denotes the Hadamard (elementwise) product of matrices.
\end{lemma}

\begin{proofof}{Proof of Lemma \ref{lem1}:}
	Since $(n-1)\mathbf{S}_n$ has a $p$-dimensional Wishart distribution with the identity covariance matrix, we get that there exists a $p \times (n-1)$ matrix $\mathbf{\tilde{Z}}$ whose entries are independent and standard normally distributed such that $(n-1)\mathbf{S}_n=\mathbf{\tilde{Z}}\mathbf{\tilde{Z}}'$. The application of Theorem 2 in \citet{bai2011asymptotic} leads to \eqref{lem1_eq1} with $\boldsymbol{\Theta}$ as in \eqref{bTheta} and $\boldsymbol{\Lambda}$ given by
	\begin{eqnarray*}
		\boldsymbol{\Lambda}&=&
		\left(
		\begin{array}{cccc}
			\lambda_{1}& \lambda_{2} & \lambda_{2}&\lambda_{2}\\
			\lambda_{2}& \lambda_{3} & \lambda_{3}&\lambda_{3}\\
			\lambda_{2}& \lambda_{3} & \lambda_{3}&\lambda_{3}\\
			\lambda_{2}& \lambda_{3} & \lambda_{3}&\lambda_{3}\\
		\end{array}
		\right)
	\end{eqnarray*}
	with
	\begin{eqnarray*}
		\lambda_{1}&=& \int_{a_{-}}^{a_{+}} z^2\text{d}F_c(z)-\left(\int_{a_{-}}^{a_{+}} z\text{d}F_c(z) \right)^2 ,\\
		\lambda_{2}&=&  1-\int_{a_{-}}^{a_{+}} z\text{d}F_c(z) \int_{a_{-}}^{a_{+}} \frac{1}{z}\text{d}F_c(z),\\
		\lambda_{3}&=& \int_{a_{-}}^{a_{+}} \frac{1}{z^2}\text{d}F_c(z)-\left(\int_{a_{-}}^{a_{+}} \frac{1}{z}\text{d}F_c(z) \right)^2\\
	\end{eqnarray*}
	where the function $F_c(z)$ denotes the cumulative distribution function of the Marchenko-Pastur law (see, \citet*{bai2010spectral}) for $c<1$ expressed as
	\begin{eqnarray*}
		\text{d}  F_c(z) = \frac{1}{2\pi z c}\sqrt{(a_{+}-z)(z-a_{-})}\mathbbm{1}_{[a_{-}, a_{+}]}(z)dz,
	\end{eqnarray*}
	where $a_{\pm}=(1\pm\sqrt{c})^2$. The moments of $F_c(z)$ present in $\boldsymbol{\Lambda}$ can be found in \citet[Lemma 14]{glombeck}. This completes the proof of the lemma.
\end{proofof}

%%%%%%%%%%%%%%%%%%%%%%%%%%%%%%%%%%%%%%%%%%%%%%%%%%%%%%%%%%%%%%%%%%%%%%%%%%%%
%%%%%%%%%%%%%%%% Lemma 2
%%%%%%%%%%%%%%%%%%%%%%%%%%%%%%%%%%%%%%%%%%%%%%%%%%%%%%%%%%%%%%%%%%%%%%%%%%%%
\begin{lemma}\label{lem2}
	Under the conditions of Theorem \ref{th1} it holds that
	\begin{equation} \begin{split}\label{lem2_eq1}
		\sqrt{n}\mathbf{h}=
		\sqrt{n}\begin{pmatrix}
		\textbf{1}_p'\hat{\mathbf{\Sigma}}_n^{-1}\mathbf{\bar{x}}_{n}-\frac{1}{1-c_n}\textbf{1}_p'\mathbf{\Sigma}^{-1}\boldsymbol{\mu}\\
		\textbf{1}_p'\hat{\mathbf{\Sigma}}_n^{-1}\mathbf{1}_p-\frac{1}{1-c_n}\textbf{1}_p'\mathbf{\Sigma}^{-1}\mathbf{1}_p\\
		\mathbf{\bar{x}}_{n}'\hat{\mathbf{\Sigma}}_n^{-1}\mathbf{\bar{x}}_{n}-\frac{1}{1-c_n}\boldsymbol{\mu}'\mathbf{\Sigma}^{-1}\boldsymbol{\mu}-\frac{c_n}{1-c_n}\\
		\mathbf{b}'\mathbf{\bar{x}}_{n}-\mathbf{b}'\boldsymbol{\mu}\\
		\mathbf{b}'\hat{\mathbf{\Sigma}}_n\mathbf{b}-\mathbf{b}'\mathbf{\Sigma}\mathbf{b}\\
		\end{pmatrix}
		\stackrel{d}{\to}
		\mathcal{N}_5\left(\mathbf{0} ,\mathbf{\Xi} \right)
		\end{split}
	\end{equation}
	
	for $c_n=p/n \to c \in[0,1)$ as $n \to \infty$ with
	\vspace{0.5cm}
\begin{equation}\label{lem2_eq2}
\mathbf{\Xi}=
\begin{pmatrix}
\frac{1}{(1-c)^3}\frac{1}{V_{GMV}}\left(s^*+\dfrac{R_{GMV}^2}{V_{GMV}}\right) & \frac{2}{(1-c)^3}\frac{R_{GMV}}{V_{GMV}^2} &\frac{2}{(1-c)^3}\frac{R_{GMV}s^*}{V_{GMV}}&\frac{1}{1-c}&-\frac{2}{1-c}R_b\\
\frac{2}{(1-c)^3}\frac{R_{GMV}}{V_{GMV}^2}& \frac{2}{(1-c)^3}\frac{1}{V_{GMV}^2}& \frac{2}{(1-c)^3}\frac{R_{GMV}^2}{V_{GMV}^2}&0&-\frac{2}{1-c}\\
\frac{2}{(1-c)^3}\frac{R_{GMV}s^*}{V_{GMV}}& \frac{2}{(1-c)^3}\frac{R_{GMV}^2}{V_{GMV}^2}&\frac{2}{(1-c)^3}\left((s^*)^2+c-1\right)&\frac{2R_b}{1-c}&-\frac{2}{1-c}R_b^2\\
\frac{1}{1-c}&0&\frac{2R_b}{1-c}&V_b&0\\
-\frac{2}{1-c}R_b&-\frac{2}{1-c}&-\frac{2}{1-c}R_b^2&0&2V_b^2\\
\end{pmatrix},
\end{equation}
	\vspace{0.3cm}
where $s^*=s+\dfrac{R_{GMV}^2}{V_{GMV}}+1$.
\end{lemma}

\begin{proofof}{Proof of Lemma \ref{lem2}:}
	Let $\mathbf{a}'=(a_1,a_2,a_3,a_4,a_5)$ be an arbitrary vector of constants. Next, we show that $\sqrt{n}\mathbf{a}'\mathbf{h}\stackrel{d}{\to}
	\mathcal{N}\left(0 , \mathbf{a}'\mathbf{\Xi}\mathbf{a} \right)$, which will prove the statement of the lemma.
	
	Since $\mathbf{x}_1,...,\mathbf{x}_n$ are independent and identically distributed with $\mathbf{x}_i \sim \mathcal{N}_p(\boldsymbol{\mu},\mathbf{\Sigma})$, we get that $\mathbf{x}_i=\boldsymbol{\mu}+\mathbf{\Sigma}\mathbf{z}_i$ where $\mathbf{z}_1,...,\mathbf{z}_n$ are independent standard normally distributed and $\mathbf{\Sigma}^{1/2}$ is the symmetric square root of $\mathbf{\Sigma}$. Moreover, it holds that
	\[\mathbf{\bar{x}}_{n}=\boldsymbol{\mu}+\mathbf{\Sigma}\mathbf{\bar{z}}_n
	\quad \mbox{and} \quad
	\hat{\mathbf{\Sigma}}_n=\mathbf{\Sigma}^{1/2}\mathbf{S}_n \mathbf{\Sigma}^{1/2},\]
	where
	\[\mathbf{\bar{z}}_{n}=\frac{1}{n}\sum_{i=1}^n\mathbf{z}_i
	\quad \mbox{and} \quad
	\mathbf{S}_n=\frac{1}{n-1}\sum_{i=1}^n(\mathbf{z}_i-\mathbf{\bar{z}}_n)(\mathbf{z}_i-\mathbf{\bar{z}}_n)'.\]
	To this end, we have that $\mathbf{\bar{z}}_{n}$ and $\mathbf{S}_n$ are independent with $\sqrt{n}\mathbf{\bar{z}}_{n}$ standard normally distributed and $(n-1)\mathbf{S}_n$ standard Wishart distributed.
	
	Let $\boldsymbol{\nu}=\mathbf{\Sigma}^{-1/2}\boldsymbol{\mu}$. We get
	\begin{eqnarray*}
		\sqrt{n}\mathbf{a}'\mathbf{h}&=&H_1(\mathbf{\bar{z}}_{n},\mathbf{S}_n)+H_2(\mathbf{\bar{z}}_{n}),
	\end{eqnarray*}
	with
	\begin{eqnarray*}
			H_1(\mathbf{\bar{z}}_{n},\mathbf{S}_n)&=& a_1\sqrt{\textbf{1}_p'\mathbf{\Sigma}^{-1}\mathbf{1}_p}\sqrt{\mathbf{\bar{x}}_{n}'\mathbf{\Sigma}^{-1}\mathbf{\bar{x}}_{n}} \sqrt{n}
		\left(\frac{\textbf{1}_p'\hat{\mathbf{\Sigma}}_n^{-1}\mathbf{\bar{x}}_{n}}
		{\sqrt{\textbf{1}_p'\mathbf{\Sigma}^{-1}\mathbf{1}_p}\sqrt{\mathbf{\bar{x}}_{n}'\mathbf{\Sigma}^{-1}\mathbf{\bar{x}}_{n}}}
		-\frac{\frac{1}{1-c_n}\textbf{1}_p'\mathbf{\Sigma}^{-1}\mathbf{\bar{x}}_{n}}
		{\sqrt{\textbf{1}_p'\mathbf{\Sigma}^{-1}\mathbf{1}_p}\sqrt{\mathbf{\bar{x}}_{n}'\mathbf{\Sigma}^{-1}\mathbf{\bar{x}}_{n}}}\right)\\
		&+&a_2\textbf{1}_p'\mathbf{\Sigma}^{-1}\mathbf{1}_p \sqrt{n}\left(\frac{\textbf{1}_p'\hat{\mathbf{\Sigma}}_n^{-1}\mathbf{1}_p}{\textbf{1}_p'\mathbf{\Sigma}^{-1}\mathbf{1}_p}-\frac{1}{1-c_n}\right)
		+a_3\mathbf{\bar{x}}_{n}'\mathbf{\Sigma}^{-1}\mathbf{\bar{x}}_{n}\sqrt{n}\left(\frac{\mathbf{\bar{x}}_{n}'\hat{\mathbf{\Sigma}}_n^{-1}\mathbf{\bar{x}}_{n}}
		{\mathbf{\bar{x}}_{n}'\mathbf{\Sigma}^{-1}\mathbf{\bar{x}}_{n}}-\frac{1}{1-c_n}\right)\\
		&+&a_5\mathbf{b}'\mathbf{\Sigma}\mathbf{b}\sqrt{n}\left(\frac{\mathbf{b}'\hat{\mathbf{\Sigma}}_n\mathbf{b}}{\mathbf{b}'\mathbf{\Sigma}\mathbf{b}}-1\right)
		=\mathbf{d}'_1(\mathbf{\bar{z}}_{n})\sqrt{n}\mathbf{h}_1(\mathbf{\bar{z}}_{n},\mathbf{S}_n)
	\end{eqnarray*}
	and
	\begin{eqnarray*}
			H_2(\mathbf{\bar{z}}_{n})&=& a_1\frac{1}{1-c_n}\sqrt{n}\left(\textbf{1}_p'\mathbf{\Sigma}^{-1}\mathbf{\bar{x}}_{n}-\textbf{1}_p'\mathbf{\Sigma}^{-1}\boldsymbol{\mu}\right)\\
		&+&a_3\frac{1}{1-c_n}\sqrt{n}\left(\mathbf{\bar{x}}_{n}'\mathbf{\Sigma}^{-1}\mathbf{\bar{x}}_{n}-\boldsymbol{\mu}'\mathbf{\Sigma}^{-1}\boldsymbol{\mu}-c_n\right)
	+a_4\sqrt{n}(\mathbf{b}'\mathbf{\bar{x}}_{n}-\mathbf{b}'\boldsymbol{\mu})\\
		%&=&a_1\frac{1}{1-c_n}\sqrt{n}\textbf{1}_p'\mathbf{\Sigma}^{-1/2}\mathbf{\bar{z}}_{n}\\
		%&+&a_3\frac{1}{1-c_n}\sqrt{n}\left(\mathbf{\bar{z}}_{n}'\mathbf{\bar{z}}_{n}+2\boldsymbol{\mu}'\mathbf{\Sigma}^{-1/2}\mathbf{\bar{z}}_{n}-c_n\right)
		%+a_4\sqrt{n}\mathbf{b}'\mathbf{\Sigma}^{1/2}\mathbf{\bar{z}}_{n}\\
		&=&\frac{a_3}{1-c_n}\sqrt{n}\left(\left(\mathbf{\bar{z}}_{n}+\mathbf{d}_2\right)'\left(\mathbf{\bar{z}}_{n}+\mathbf{d}_2\right)-\mathbf{d}_2'\mathbf{d}_2-c_n\right)	
	\end{eqnarray*}
	with
	\begin{eqnarray*}
		\mathbf{d}_1(\mathbf{\bar{z}}_{n})=\begin{pmatrix}
			a_5\mathbf{b}'\mathbf{\Sigma}\mathbf{b}\\
			a_2\textbf{1}_p'\mathbf{\Sigma}^{-1}\mathbf{1}_p \\
			a_1\sqrt{\textbf{1}_p'\mathbf{\Sigma}^{-1}\mathbf{1}_p}\sqrt{(\boldsymbol{\nu}+\mathbf{\bar{z}}_{n})'(\boldsymbol{\nu}+\mathbf{\bar{z}}_{n})} \\
			a_3(\boldsymbol{\nu}+\mathbf{\bar{z}}_{n})'(\boldsymbol{\nu}+\mathbf{\bar{z}}_{n})
		\end{pmatrix},
	\end{eqnarray*}
		\begin{eqnarray*}
			\mathbf{d}_2=\frac{1-c_n}{a_3}\left(\frac{a_3}{1-c_n}\mathbf{\Sigma}^{-1/2}\boldsymbol{\mu}+\frac{a_1}{2(1-c_n)}\mathbf{\Sigma}^{-1/2}\textbf{1}_p
			+\frac{a_4}{2}\mathbf{\Sigma}^{1/2}\mathbf{b}\right),
	\end{eqnarray*}
	and
\begin{eqnarray*}
	\mathbf{h}_1(\mathbf{\bar{z}}_{n},\mathbf{S}_n)=
	\begin{pmatrix}
		\frac{\mathbf{b}'\mathbf{\Sigma}^{1/2}\mathbf{S}_n\mathbf{\Sigma}^{1/2}\mathbf{b}}{\mathbf{b}'\mathbf{\Sigma}\mathbf{b}}-1\\
		\frac{\textbf{1}_p'\mathbf{\Sigma}^{-1/2}\mathbf{S}_n^{-1}\mathbf{\Sigma}^{-1/2}\mathbf{1}_p}{\textbf{1}_p'\mathbf{\Sigma}^{-1}\mathbf{1}_p}-\frac{1}{1-c_n}\\
		\frac{\textbf{1}_p'\mathbf{\Sigma}^{-1/2}\mathbf{S}_n^{-1}(\boldsymbol{\nu}+\mathbf{\bar{z}}_{n})}
		{\sqrt{\textbf{1}_p'\mathbf{\Sigma}^{-1}\mathbf{1}_p}\sqrt{(\boldsymbol{\nu}+\mathbf{\bar{z}}_{n})'(\boldsymbol{\nu}+\mathbf{\bar{z}}_{n})}}
		-\frac{1}{1-c_n}\frac{\textbf{1}_p'\mathbf{\Sigma}^{-1/2}(\boldsymbol{\nu}+\mathbf{\bar{z}}_{n})}
		{\sqrt{\mathbf{1}_p'\mathbf{\Sigma}^{-1}\mathbf{1}_p}\sqrt{(\boldsymbol{\nu}+\mathbf{\bar{z}}_{n})'(\boldsymbol{\nu}+\mathbf{\bar{z}}_{n})}}\\
		\frac{(\boldsymbol{\nu}+\mathbf{\bar{z}}_{n})'\mathbf{S}_n^{-1}(\boldsymbol{\nu}+\mathbf{\bar{z}}_{n})}
		{(\boldsymbol{\nu}+\mathbf{\bar{z}}_{n})'(\boldsymbol{\nu}+\mathbf{\bar{z}}_{n})}-\frac{1}{1-c_n}
	\end{pmatrix}.
\end{eqnarray*}
	
	Since $\mathbf{S}_n$ and $\mathbf{\bar{z}}_{n}$ are independent the conditional distribution of $H_1(\mathbf{\bar{z}}_{n},\mathbf{S}_n)$ given $\mathbf{\bar{z}}_{n}=\mathbf{v}$ coincides with $H_1(\mathbf{v},\mathbf{S}_n)$. Furthermore, the application of Lemma \ref{lem1} to $\sqrt{n}\mathbf{h}_1(\mathbf{v},\mathbf{S}_n)$ proves that it is asymptotically normally distributed and, thus, the asymptotic stochastic representation of $H_1(\mathbf{\bar{z}}_{n},\mathbf{S}_n)$ is given by
	\begin{equation}\label{H1zS}
		H_1(\mathbf{\bar{z}}_{n}\mathbf{v},\mathbf{S}_n) \stackrel{d}{=} \sqrt{\frac{2}{c}}\,
		\sqrt{\mathbf{d}'_1\left(\boldsymbol{\Theta}\left(\frac{\mathbf{\Sigma}^{1/2}\mathbf{b}}{\sqrt{\mathbf{b}'\mathbf{\Sigma}\mathbf{b}}},
			\frac{\mathbf{\Sigma}^{-1/2}\mathbf{1}_p}{\sqrt{\textbf{1}_p'\mathbf{\Sigma}^{-1}\mathbf{1}_p}},
			\frac{(\boldsymbol{\nu}+\mathbf{\bar{z}}_{n})}{\sqrt{(\boldsymbol{\nu}+\mathbf{\bar{z}}_{n})'(\boldsymbol{\nu}+\mathbf{\bar{z}}_{n})}}\right)
			\circ\boldsymbol{\Lambda} \right) \mathbf{d}_1}\omega_1 ,
	\end{equation}
	
	where $\omega_1 \stackrel{d}{\to} \mathcal{N}(0,1)$ and is independent of $\mathbf{\bar{z}}_{n}$ and hence of $H_2(\mathbf{\bar{z}}_{n})$. Finally, we have that $n\left(\mathbf{\bar{z}}_{n}+\mathbf{d}_2\right)'\left(\mathbf{\bar{z}}_{n}+\mathbf{d}_2\right)$ has a non-central $\chi^2$ distribution with $p$ degrees of freedom and noncentrality parameter $n\mathbf{d}_2'\mathbf{d}_2$. The application of \citet[]{Bodnar2016125} leads to
	
 \begin{equation*}
		\sqrt{p}\left(\frac{n\left(\mathbf{\bar{z}}_{n}+\mathbf{d}_2\right)'\left(\mathbf{\bar{z}}_{n}+\mathbf{d}_2\right)}{p}-\frac{n\mathbf{d}_2'\mathbf{d}_2}{p}-1\right)
		\stackrel{d}{\to} \mathcal{N}\left(0,2+4\frac{\mathbf{d}_2'\mathbf{d}_2}{c}\right)
		\end{equation*}
	and, consequently,
	\begin{equation}\label{H2z}
	H_2(\mathbf{\bar{z}}_{n}) \stackrel{d}{=}\frac{\sqrt{c_n}}{1-c_n}a_3 \sqrt{2+4\frac{\mathbf{d}_2'\mathbf{d}_2}{c_n}} \omega_2.
	\end{equation}
	where $\omega_2 \stackrel{d}{\to} \mathcal{N}(0,1)$.
	
	Using that $\boldsymbol{\nu}'\mathbf{\bar{z}}_{n}  \stackrel{a.s.}{\to} 0$ and $\mathbf{\bar{z}}_{n}'\mathbf{\bar{z}}_{n}  \stackrel{a.s.}{\to} c$, the application of Slutsky's lemma (c.f., \citet[Theorem 1.5]{dasgupta2008}) leads to
	\[
	\sqrt{n}\mathbf{a}'\mathbf{h}
	\stackrel{d}{\to} \mathcal{N}(0,\mathbf{a}' \mathbf{\Xi}\mathbf{a})
	\]
	for $p/n \to c+o(n^{-1/2})$ as $n \to \infty$ where $\mathbf{\Xi}$ is given in \eqref{lem2_eq2}. Since $\mathbf{a}$ is an arbitrary vector, the statement of Lemma \ref{lem2} is proved.
\end{proofof}

\begin{proofof}{Proof of Theorem \ref{th1}:}
	It holds that
	\begin{eqnarray*}
			\hat{R}_{GMV}-R_{GMV}&=&
			\hat{V}_{GMV}\Biggl(\left(\textbf{1}_p'\hat{\mathbf{\Sigma}}_n^{-1}\mathbf{\bar{x}}_{n}-\frac{1}{1-c_n}\textbf{1}_p'\mathbf{\Sigma}^{-1}\boldsymbol{\mu}\right)-
			R_{GMV}\left(\textbf{1}_p'\hat{\mathbf{\Sigma}}_n^{-1}\textbf{1}_p-\frac{1}{1-c_n}\textbf{1}_p'\mathbf{\Sigma}^{-1}\textbf{1}_p\right)\Biggr),\\	
			\hat{V}_{c}-V_{GMV}&=&-V_{GMV}\hat{V}_{GMV}\left(\textbf{1}_p'\hat{\mathbf{\Sigma}}_n^{-1}\textbf{1}_p-\frac{1}{1-c_n}\textbf{1}_p'\mathbf{\Sigma}^{-1}\textbf{1}_p\right)
	\end{eqnarray*}
	and
	{\small \begin{eqnarray*}
			\hat{s}_{c}-s_{GMV}%&=&(1-c_n)\hat{s}-s-c_n\\
			&=& (1-c_n)\left(\mathbf{\bar{x}}_{n}'\hat{\mathbf{\Sigma}}_n^{-1}\mathbf{\bar{x}}_{n}-\frac{1}{1-c_n}\boldsymbol{\mu}'\mathbf{\Sigma}^{-1}\boldsymbol{\mu}-\frac{c_n}{1-c_n}\right)\\
			&-&(1-c_n)\left(\frac{(\textbf{1}_p'\hat{\mathbf{\Sigma}}_n^{-1}\mathbf{\bar{x}}_{n})^2}{\textbf{1}_p'\hat{\mathbf{\Sigma}}_n^{-1}\textbf{1}_p}
			-\frac{1}{1-c_n}\frac{(\textbf{1}_p'\mathbf{\Sigma}^{-1}\boldsymbol{\mu})^2}{\textbf{1}_p'\mathbf{\Sigma}^{-1}\textbf{1}_p}\right)\\
			&=&(1-c_n)\left(\mathbf{\bar{x}}_{n}'\hat{\mathbf{\Sigma}}_n^{-1}\mathbf{\bar{x}}_{n}-\frac{1}{1-c_n}\boldsymbol{\mu}'\mathbf{\Sigma}^{-1}\boldsymbol{\mu}-\frac{c_n}{1-c_n}\right)\\
			&-&(1-c_n)\hat{V}_{GMV}\Bigg(
			\left(\textbf{1}_p'\hat{\mathbf{\Sigma}}_n^{-1}\mathbf{\bar{x}}_{n}+\frac{1}{1-c_n}\frac{R_{GMV}}{V_{GMV}}\right)
			\left(\textbf{1}_p'\hat{\mathbf{\Sigma}}_n^{-1}\mathbf{\bar{x}}_{n}-\frac{1}{1-c_n}\textbf{1}_p'\mathbf{\Sigma}^{-1}\boldsymbol{\mu}\right)\\
			&-&\frac{1}{1-c_n}\frac{R_{GMV}^2}{V_{GMV}}
			\left(\textbf{1}_p'\hat{\mathbf{\Sigma}}_n^{-1}\textbf{1}_p-\frac{1}{1-c_n}\textbf{1}_p'\mathbf{\Sigma}^{-1}\textbf{1}_p\right)\Bigg).
	\end{eqnarray*}}
	
	Hence,
	\[
	\sqrt{n}\begin{pmatrix}
	\hat{R}_{GMV}- R_{GMV}\\
	\hat{V}_{c}-V_{GMV}\\
	\hat{s}_c-s\\
	\hat{R}_b-R_b\\
	\hat{V}_{b}-V_b
	\end{pmatrix}=\mathbf{D}\sqrt{n}\mathbf{h},
	\]
	with $\mathbf{h}$ is defined in \eqref{lem2_eq1} and
	\begin{equation*}
		\mathbf{D}=
		\begin{pmatrix}
		(1-c_n)\hat{V}_{c}&-(1-c_n)\hat{V}_{c}R_{GMV}&0&0&0\\
		0&-(1-c_n)\hat{V}_{c}V_{GMV}&0&0&0\\
		(1-c_n)\hat{V}_{c}\left(\frac{R_{GMV}}{V_{GMV}}-\frac{\hat{R}_{GMV}}{\hat{V}_{c}}\right)&(1-c_n)\hat{V}_{c}\frac{R_{GMV}^2}{V_{GMV}}&(1-c_n)&0&0\\
		0&0&0&1&0\\
		0&0&0&0&1\\
		\end{pmatrix}
	\end{equation*}
	
	The application of $\hat{R}_{GMV}\stackrel{a.s.}{\to} R_{GMV}$ and $\hat{V}_{c}\stackrel{a.s.}{\to}V_{GMV}$ for $p/n \to c \in [0,1)$ as $n \to \infty$, the results of Lemma \ref{lem2}, and Slutsky's lemma (c.f., \citet[Theorem 1.5]{dasgupta2008}) completes the proof of the theorem.
\end{proofof}

%\begin{proofof}{Proof of Theorem \ref{th4}:}
%to be done
%\end{proofof}

\end{document}